\documentclass{article}

\PassOptionsToPackage{numbers, compress}{natbib}
\usepackage[final]{neurips_2022}

\usepackage[utf8]{inputenc} % allow utf-8 input
\usepackage[T1]{fontenc}    % use 8-bit T1 fonts
\usepackage{hyperref}       % hyperlinks
\usepackage{url}            % simple URL typesetting
\usepackage{booktabs}       % professional-quality tables
\usepackage{amsfonts}       % blackboard math symbols
\usepackage{amsmath}
\usepackage{amssymb}
\usepackage{mathtools}
\usepackage{amsthm}
\usepackage{nicefrac}       % compact symbols for 1/2, etc.
\usepackage{microtype}      % microtypography
\usepackage{xcolor}         % colors

\usepackage{bm}
\usepackage{caption}
\usepackage{subcaption}
\usepackage{hyperref}
\usepackage{multirow}

\usepackage{color}
\usepackage[group-separator={,},group-minimum-digits={3}]{siunitx}
\usepackage{enumitem}

\makeatletter
\renewcommand*\subsectionautorefname{\S\@gobble}
\renewcommand*\sectionautorefname{\S\@gobble}
\makeatother

\title{Museformer: Transformer with Fine- and Coarse-Grained Attention for Music Generation}

\author{%
	{Botao Yu$^\dagger$, \quad Peiling Lu$^\ddagger$, \quad Rui Wang$^\ddagger$, \quad Wei Hu$^\dagger$\thanks{Wei Hu and Xu Tan are the corresponding authors. This work was partially done while the first author was interning at Microsoft Research Asia.}\ \ , \quad Xu Tan$^{\ddagger\,*}$,}\\
	{\textbf{Wei Ye$^\S$, \quad Shikun Zhang$^\S$, \quad Tao Qin$^\ddagger$, \quad Tie-Yan Liu$^\ddagger$}}\\
	$^\dagger$State Key Laboratory for Novel Software Technology, Nanjing University, China\\ 
	$^\ddagger$Microsoft Research Asia\\
	$^\S$National Engineering Research Center for Software Engineering, Peking University, China\\
	\texttt{btyu@foxmail.com, \{peil,ruiwa,xuta,taoqin,tyliu\}@microsoft.com,}\\
	\texttt{whu@nju.edu.cn, \{wye,zhangsk\}@pku.edu.cn} \\ \\
	\url{https://github.com/microsoft/muzic}
}

\begin{document}

\maketitle

\setcounter{footnote}{0}

\begin{abstract}

Symbolic music generation aims to generate music scores automatically.
A recent trend is to use Transformer or its variants in music generation, which is, however, suboptimal, because the full attention cannot efficiently model the typically long music sequences (e.g., over \num{10000} tokens), and the existing models have shortcomings in generating musical repetition structures.
In this paper, we propose Museformer, a Transformer with a novel fine- and coarse-grained attention for music generation. 
Specifically, with the fine-grained attention, a token of a specific bar directly attends to all the tokens of the bars that are most relevant to music structures (e.g., the previous 1st, 2nd, 4th and 8th bars, selected via similarity statistics); 
with the coarse-grained attention, a token only attends to the summarization of the other bars rather than each token of them so as to reduce the computational cost. 
The advantages are two-fold. 
First, it can capture both music structure-related correlations via the fine-grained attention, and other contextual information via the coarse-grained attention. 
Second, it is efficient and can model over $3\times$ longer music sequences compared to its full-attention counterpart. 
Both objective and subjective experimental results demonstrate its ability to generate long music sequences with high quality and better structures.\footnote{The generated music samples can be found at \url{https://ai-muzic.github.io/museformer}. The source code can be found at \url{https://github.com/microsoft/muzic}.}
\end{abstract}

\section{Introduction}
	\label{sec:intro}
	
\begin{figure}
		\centering
		\includegraphics[width=\textwidth]{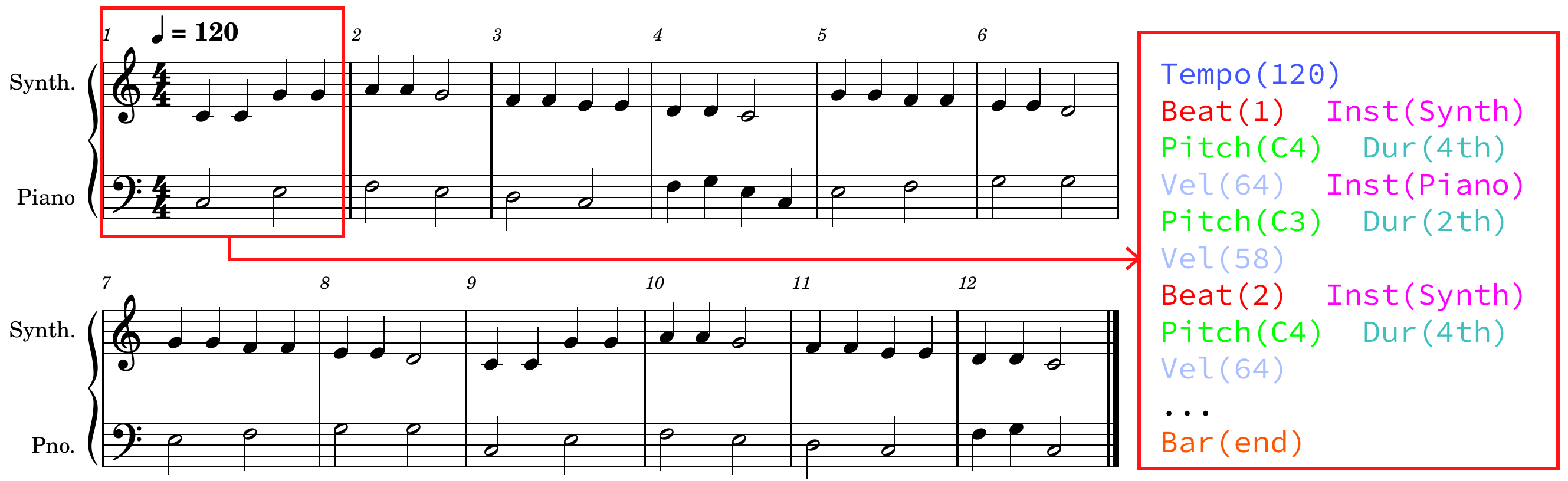}
		\caption{The music score of \textit{Twinkle, Twinkle, Little Star} and its corresponding token representation. Every two consecutive bars on the \textit{Synth.} track have the same rhythmic pattern, and the 9th - 12th bars repeat the 1st - 4th bars with an interval of 8 bars. Structures embodied as repetitions and variations are common in music.}
		\label{fig:score}
	\end{figure}

	Symbolic music generation aims at generating music scores automatically and has drawn more and more attention in recent years \cite{garcia2011automatic,yanchenko2017classical,huang2018music}. Since music can be represented in organized sequences of discrete tokens just like text, Transformer-based models, which have been demonstrated to work well on text generation \cite{radford2018improving,ijcai2021p612}, are increasingly applied in music generation \cite{huang2018music,donahue2019lakhnes,muhamed2021symbolic,roberts2018hierarchical,ju2021telemelody,ren2020popmag,sheng2021songmass} and have made great success. 
	% While the self-attention mechanism empowers Transformer to directly capture the correlations over the whole input sequence, the quadratic complexity limits its scalability to music sequences especially for the multi-instrument polyphonic music, as they are typically very long, e.g., over \num{10000} tokens.
	While the self-attention mechanism empowers Transformer to capture the complex correlations in music, there are two ubiquitous challenges to solve for this task: 1) \emph{Long sequence modeling}. Music sequences are typically very long, especially for multi-instrument polyphonic music where the lengths can usually exceed \num{10000}. The quadratic complexity of full attention limits its scalability to that length. 2) \emph{Music structure modeling}. Music has its unique structures, where a piece can usually repeat some patterns of a previous piece, occasionally with some variations, after either a short or a long distance (see \autoref{fig:score} for an example). Successfully generating reasonable structures would make the music more realistic just like human-made music.

	Although many Transformer variants in natural language processing (NLP) \cite{lin2021survey} have been proposed to handle long sequences (the first challenge), they cannot well model the music structures (the second challenge). 
	According to the basic principles of these models, we can classify them into two types. The first type is called \textit{local focusing}. Models of this type, e.g., Transformer-XL \cite{dai2019xl} and Longformer \cite{beltagy2020longformer}, mainly focus on part of the input sequence, and drop the rest tokens to reduce the cost. However, the parts that they focus on cannot contain many of the essential ranges important to music structures, and directly dropping the rest tokens may lead to losing some important information. The second type is called \textit{global approximation}. Models of this type such as Linear Transformer \cite{katharopoulos2020transformers} utilize linearized attention or sequence compression over the whole input sequence to approximate the token pair-wise attention. While the approximation effectively reduces the complexity, they cannot accurately capture the correlations between related parts and accordingly are inadequate in generating repetition structures. We will review more existing long-sequence Transformers in \autoref{sec:related_work}. Directly applying these models to music generation is suboptimal, and it is desirable to design an efficient model that can well model long music sequences as well as their structures.

	In this paper, we propose to unify the above two types of models, which can well fit the characteristics of music. Our motivation is based on the observation that the importance is not uniformly distributed over the music sequence, and thus we do not need to treat all the tokens equally.
	Intuitively, to generate music with repetition structures, the most important information the model should directly refer to when generating a music bar, lies in those bars that tend to be repeated in the current bar. We call these bars \textit{structure-related bars}. For the other bars that are less important, approximation should do the trick. To this end, the proposed fine- and coarse-grained attention takes two different schemes towards different bars: with the fine-grained attention, tokens of the structure-related bars are directly attended to, to well learn the structure-related correlations; with the coarse-grained attention, information of the other bars are captured via summarization \cite{ye2019bp}, rather than attending to each token of them, to decrease the computation and space complexity, and meanwhile retain the necessary information of these bars. 
	The structure-related bars are selected according to the statistics on human-made music. They are not necessarily the most recent and consecutive bars but can include distant ones, to cater to long-term structures in music. 
	Our main contributions in this paper are summarized as follows:
	\begin{itemize}[leftmargin=*]
		\item We propose Museformer, a Transformer model with a novel fine- and coarse-grained attention for music generation. It captures the correlations of structure-related bars via fine-grained attention for learning the music structures, as well as the necessary and concentrated information of the other bars through their summarization via coarse-grained attention.
		\item We propose to select the structure-related bars based on similarity statistics on human-made music, which can help decrease the perplexity and generate music that exhibits better structures.
		\item The computation complexity and space complexity are reduced greatly to nearly linear in practice, which enables Museformer to scale up to long music sequences.
		\item Experimental results show that Museformer can generate music of full-song length with high quality and better structures.
	\end{itemize}

\section{Related Work}
\label{sec:related_work}
	
	\subsection{Symbolic Music Generation}
	
	Symbolic music generation aims to exploit machines to compose music scores automatically. It has been attracting more and more people to work on it, and the solutions continuously evolve from rule-based models \cite{garcia2011automatic} to probabilistic models \cite{farbood2001analysis,allan2002harmonising,yanchenko2017classical} to deep learning models \cite{donahue2019lakhnes,muhamed2021symbolic,roberts2018hierarchical,ju2021telemelody,ren2020popmag,sheng2021songmass}.  
	
	In recent years, the Transformer-based models have achieved great success in many text generation tasks \cite{radford2018improving,ijcai2021p612}, thus have also been increasingly applied in the similar music generation tasks. \citet{huang2018music} apply Transformer in symbolic music generation for the first time and show that it can achieve better performance compared to previous deep learning models such as recurrent neural networks. While Transformer shows promising results in music generation, the quadratic complexity of the attention mechanism limits its applications on the typically long music sequences. To resolve this problem, researchers on symbolic music generation come up with different solutions. One popular solution is to design a new representation method to represent musical information in fewer tokens, such as compound word \cite{hsiao2021compound} and OctupleMIDI \cite{musicbert}. While this solution indeed decreases the lengths of music sequences to a certain extent, full-song music sequences are still too long for Transformer to handle. Another solution is to use a long-sequence Transformer variant as the backbone. For example, \citet{huang2020pop} use Transformer-XL \cite{dai2019xl}, and \citet{hsiao2021compound} use Linear Transformer \cite{katharopoulos2020transformers}. Although these models can process long sequences, they are initially designed for NLP tasks and cannot well model the music structures. 
	% In view of the fact that music has its unique structures different from text, directly applying them in music generation should be suboptimal, as we have mentioned in Introduction and will give a more extensive discussion in the following section. The existing Transformer-based models for music generation directly borrow the long-sequence Transformers from NLP to generate long music sequence, without considering music's characteristics. To the best of our knowledge, we are the first to propose a Transformer variant that are targeted at symbolic music generation.
	We will introduce and discuss more about the long-sequence Transformers in the following section.

	\subsection{Long-Sequence Transformers}
	\label{subsec:ls_tfm}
	
	Many Transformer variants have been proposed to tackle long-sequence tasks \cite{lin2021survey}, which in general can be categorized into the following types. 
	1) Recurrent Transformer, which encodes sequences chunk by chunk \cite{dai2019xl,rae2019compressive,ding2021ernie}. 
	2) Sparse attention, which sparsifies the attention layout with either predefined patterns \cite{beltagy2020longformer,child2019generating,zaheer2020big,ye2019bp,parmar2018imagetfm}, or with content-based patterns \cite{Kitaev2020Reformer,routingtfm,sac,ssa,wu2021smart}. 
	3) Linearized attention, which replaces the exponent of the inner product of features with the multiplication of feature maps  \cite{katharopoulos2020transformers,choromanski2020masked, choromanski2020rethinking,peng2021random}. 
	4) Compression-based attention, which reduces the number of queries or key-value pairs by compressing the contextual representations \cite{wang2020linformer,poolingformer,cluster}.
	In addition, some models also try to combine multiple techniques, such as Compressive Transformer \cite{rae2019compressive} that combines recurrent and compression-based methods, Poolingformer \cite{poolingformer} and Transformer-LS \cite{chen2021tfmls} that combine sparse attention and compression-based methods.
	
	Existing works on music generation directly adopt some of those long-sequence Transformers to process long music sequences, but it is suboptimal due to the unique structures of music.
	In general, music has many repeating or similar pieces, many of which are distant from each other, and the distance is measured by time units such as bar or beat instead of the number of tokens.
	Therefore, the receptive fields of the existing recurrent Transformers or sparse attention methods cannot cover the many ranges of structure-related content. Although the linearized or compression-based attention can cover the whole sequence, they do not precisely capture the correlation between each pair of tokens and may have shortcomings in generating repeating or similar music pieces.
	% To the best of our knowledge, Museformer is the first Transformer model that is targeted at solving the long music structure modeling problem.

%	Variant transformer-based models are widely used for modeling long sequence in NLP by reducing quadratic complexity of self-attention module. \citet{beltagy2020longformer} propose an attention mechanism that scales linearly with sequence length by combining a windowed local-context self-attention and an end task motivated global attention. \citet{katharopoulos2020transformers} develop a kernel-based formulation of self-attention for improving the method of calculating attention weights. \citet{dai2019xl} introduces a segment-level recurrence mechanism and a novel positional encoding scheme.
%	
%	Variety of approaches are proposed to handle long sequences for NLP, image and speech tasks. Since it is observed that the attention matrix is usually sparse \cite{child2019generating}, some works consider to only compute selected query-key pairs to reduce the complexity of attention, where the query-key pairs may be selected based on the position \cite{beltagy2020longformer} or the input content \cite{roy2021efficient}. Based on approximation algorithms, the computational complexity of the attention can be significantly reduced \cite{katharopoulos2020transformers,wang2020linformer}, which unfortunately affects the accuracy of the attention matrix.

\section{Museformer}
    
    In order to well model long music sequences as well as their music structures, we propose Museformer. 
    The general idea is that we do not need to focus on the whole sequence with the same importance level given that the complexity of pair-wise full attention is unacceptably high, but instead we combine two different attention schemes -- fine-grained attention for the structure-related bars, and coarse-grained attention for the other bars. 
    Museformer basically follows the original Transformer architecture \cite{vaswani2017attention}, and a novel fine- and coarse-grained attention (FC-Attention) is designed to replace the original self-attention module to tackle the challenges in long music sequence modeling.
	
	Museformer takes a music token sequence $X = X_1, \ldots, X_{b}$ as input, where $b$ denotes the number of music bars, and the $i$-th bar $X_i= x_{i,1}, \ldots, x_{i, |X_i|} $ contains $|X_i|$ tokens. For the $i$-th bar where $i=1,\ldots, b$, we insert a \textit{summary token} $s_i$ after it to facilitate local information aggregation in FC-Attention. The summary token sequence is denoted by $S=s_1, \ldots, s_b$. After the insertion, the overall token sequence becomes $X_1, s_1 \cdots, X_b, s_b$.
	The embedding layer embeds the overall token sequence into a vector space, and concatenates bar embeddings and beat embeddings as the positional information \cite{ren2020popmag}, followed by a linear projection. 
	The Museformer layers then model the contextual representation, and the hidden state output by the last layer is fed into a softmax classifier to predict the next token. 
	
	In the following part, we first give a brief formulation of the attention mechanism (\autoref{subsec:preliminary}). Then, we introduce the details of FC-Attention (\autoref{subsec:attn}), followed by the selection of the structure-related bars (\autoref{subsec:bar_selection}). Finally, we discuss the merits of Museformer (\autoref{subsec:analysis}).
	
	\subsection{Preliminary: Attention}
	\label{subsec:preliminary}
	
	The attention mechanism \cite{vaswani2017attention} is the basis of FC-Attention. It receives two sequential inputs, namely source $\bm{X} \in \mathbb{R}^{n_\text{src} \times d}$ and target $\bm{X}' \in \mathbb{R}^{n_\text{tgt} \times d}$, where $n_\text{src}$ and $n_\text{tgt}$ are the sequence lengths of source and target respectively, $d$ is the embedding dimension. It computes contextual representation for each $\bm{x}'_i \in \mathbb{R}^{1\times d}$ in target $\bm{X}'$ as
	\begin{equation}
		\text{Attn}(\bm{x}'_i, \bm{X}) = \text{softmax}\Big(\frac{\bm{x}'_i\bm{W}_Q   {(\bm{X}\bm{W}_K)}^T}{\sqrt{d}}\Big) \bm{X}\bm{W}_V,
		\label{eq:attn}
	\end{equation}
	% Here, we formulate attention with respect to a single item $\bm{x}_i$, because in our model, we assign different $\bm{X}'$ for different $\bm{x}_i \in X$. 
	where $\bm{W}_Q, \bm{W}_K, \bm{W}_V \in \mathbb{R}^{d \times d}$ are the trainable parameters.
	In practice, we employ the multi-head version of attention, but omit it from the equation for simplicity.
	
	% Combined with the first paragraph of this section.
	% \paragraph{Token Notations}
	% Museformer receives a music token sequence as the input, denoted by $X = (X_1, \ldots, X_{b})$, where $b$ is the number of music bars, and the $i$-th bar $X_i=( x_{i,1}, \ldots, x_{i, |X_i|} )$ contains $|X_i|$ tokens. 
	% We then insert a summary token after each bar to summarize its local information in the FC-Attention computation. The summary token for the $i$-th bar is denoted by $s_i$, and the summary token sequence is denoted as $S=( s_1, \ldots, s_b )$.
	
	\subsection{Fine- and Coarse-Grained Attention}
	\label{subsec:attn}
	
\begin{figure}
		\centering
		
		\begin{subfigure}[t]{0.60\textwidth}
			\centering
			\includegraphics[width=\textwidth]{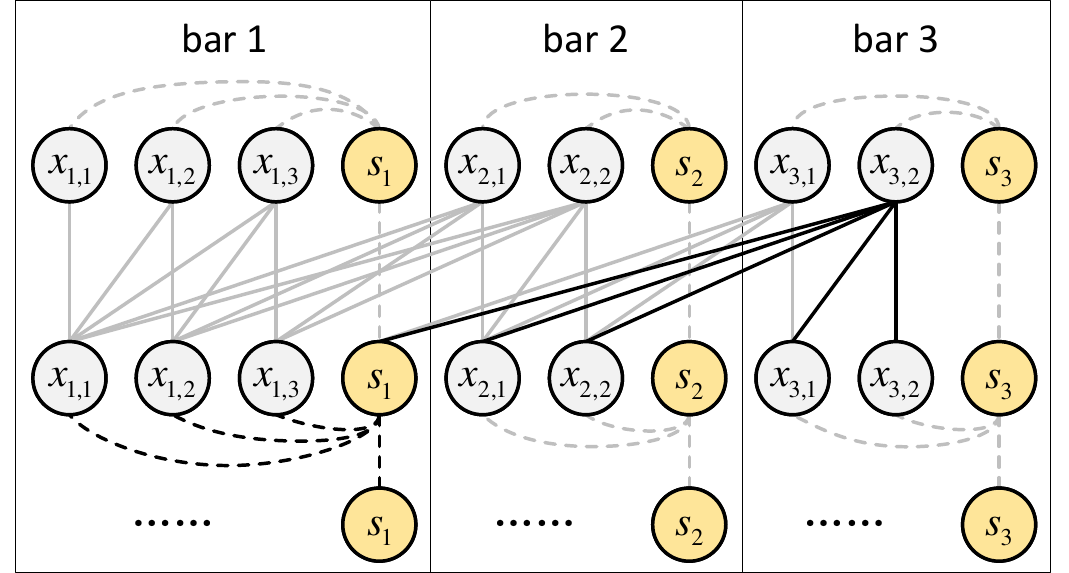}
			\caption{Information flow. The dashed/solid lines depict the summarization/aggregation step. The dark lines represent the information flow towards $x_{3,2}$.}
			\label{subfig:fc_attn_layer}
		\end{subfigure}
		\hfill
		\begin{subfigure}[t]{0.335\textwidth}
			\centering
			\includegraphics[width=\textwidth]{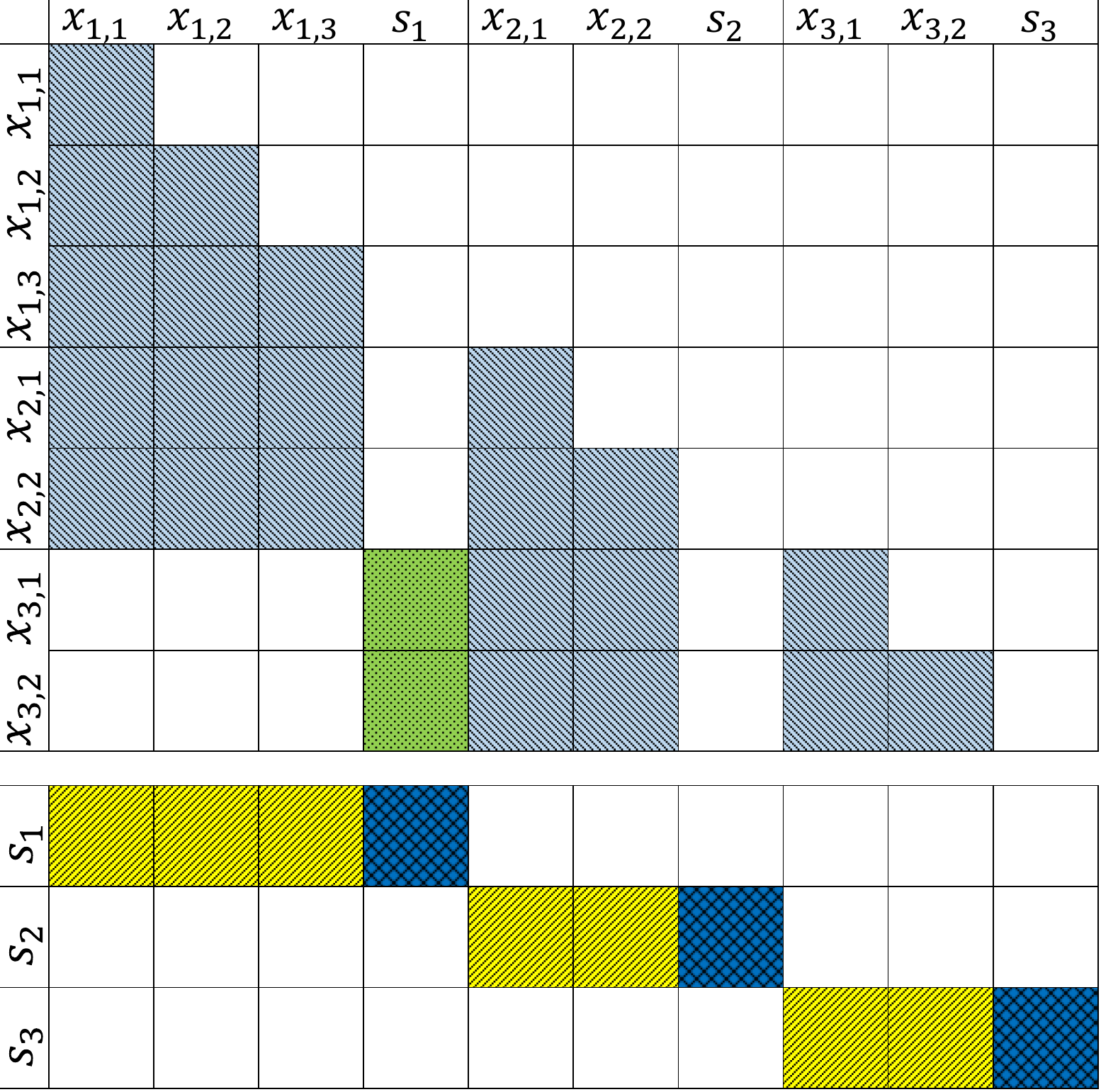}
			\caption{Attention layouts. The bottom/top part corresponds to the summarization/aggregation step.}
			\label{subfig:fc_attn_matrix}
		\end{subfigure}
		
		\caption{The fine- and coarse-grained attention. It shows a toy example of 3 bars, where for each bar, only the previous 1st bar is regarded as the structure-related bar.}
		\label{fig:fc_attn}
	\end{figure}

	The basic idea of FC-Attention is that, instead of directly attending to all the tokens which causes the quadratic complexity, a token of a specific bar only directly attends to the structure-related bars that are essential for generating structured music (fine-grained attention), and for the other bars, the token only attends to their summary tokens to obtain concentrated information (coarse-grained attention). To achieve this, we first summarize the local information of each bar through the \textit{summarization} step, and then aggregate the fine-grained and coarse-grained information through the \textit{aggregation} step. \autoref{fig:fc_attn} visualizes the process.
	
	\paragraph{Summarization}
	In the summarization step, we aggregate the information of each bar into the corresponding summary token. For the $i$-th bar, given the representation of summary token $\bm{s}_i \in \mathbb{R}^{1 \times d}$, and that of the music tokens $\bm{X}_i = [\bm{x}_{i,1}, \ldots, \bm{x}_{i,|X_i|}] \in \mathbb{R}^{|X_i| \times d}$, the summarization of this bar is 
	\begin{equation}
		\tilde{\bm{s}}_i = \text{Attn} \big(  \bm{s}_i, [  \bm{X}_i, \bm{s}_i  ] \big),
		\label{eq:summarization}
	\end{equation}
	where $\text{Attn}(\cdot)$ is defined in \autoref{eq:attn}, $[\cdot]$ is the concatenation operation. In this step, each summary token attends to music tokens of its corresponding bar as well as the summary token itself.
	
	\paragraph{Aggregation}
	In the aggregation step, we aggregate the information of the tokens belonging to the structure-related bars or within the current bar, as well as the summarization of the other bars, so as to update the contextual representations of music tokens. The updated representation for $x_{i,j}$ is
	\begin{equation}
		\tilde{\bm{x}}_{i,j} = \text{Attn}\Big(  
			\bm{x}_{i,j},
			[    \bm{X}_{R(i)}, \bm{X}_{i, k\le j}  ,   \bm{\tilde{S}}_{\bar{R}(i)}  ]  \Big),
		\label{eq:aggregation}
	\end{equation}
	where $R(i)$ is the set of the indices of the structure-related bars with respect to the $i$-th bar, and $\bar{R}(i)$ is the set of the indices of other previous bars. $\bm{X}_{R(i)}$ is the matrix formed by stacking $\{\mathbf{X}_i \, | \, i \in R(i)\}$. Similarly, $\bm{\tilde{S}}_{\bar{R}(i)}$ is the matrix formed by stacking $\{\tilde{\bm{s}}_i \, | \, i \in \bar{R}(i)\}$. $\bm{X}_{i, k\le j}$ is the matrix formed by stacking $\{\bm{x}_{i,k} \,|\, \bm{x}_{i,k} \in \bm{X}_i, \, k \le j \}$. In other words, the music token $x_{i,j}$ only attends to those music tokens that belong to the structure-related bars, and its previous tokens within the current bar. For the other bars, it only attends to their summary tokens.
	
	% Made this part as a standalone subsection
	% [* The description of this subsection is essential and need to refine several times. *]
	\subsection{Structure-Related Bar Selection}
	\label{subsec:bar_selection}
	% We conduct a statistical analysis on the POP909 dataset\cite{wang2020POP909}, where the segments of a song are annotated, through which we can easily calculate the distribution of the repetition intervals, i.e., the distance between two repeating bars. As shown in \autoref{fig:repetition_statistics}, music tends to repeat every 4 bars or its multipliers in most cases, and this also complies with our musical expertise and experience. 
	Considering that the structure-related bars are expected to contain those bars that are most likely to be repeated by the current bar to be generated, we propose to pinpoint them by similarity statistics, and the bars with high similarities should be selected.
	Specifically, for each song in the training set, we calculate the similarity between each pair of the bars, which is defined as
	\begin{equation}
	    l_{i,j} = \frac{ | N(i) \cap N(j) | }{ | N(i) \cup N(j) | },
	\end{equation}
	where $N(i)$ is the set of music notes within the $i$-th bar, and two notes are considered equal when their pitches, duration, and onset positions within their bars are all the same. The value of $l_{i,j}$ ranges from $0.0$ to $1.0$. If two bars are exactly the same, this value equals $1.0$.
	We then calculate the average similarity of those bar pairs whose intervals are $t$ over the training set $D$, which is formulated as
	\begin{equation}
	    \label{eq:similarity_distribution}
	    L_{t} = \text{Mean}( \sum_D \sum_{j=i+t} l_{i,j} ).
	\end{equation}
	
\begin{figure}
		\centering
		\includegraphics[width=\columnwidth]{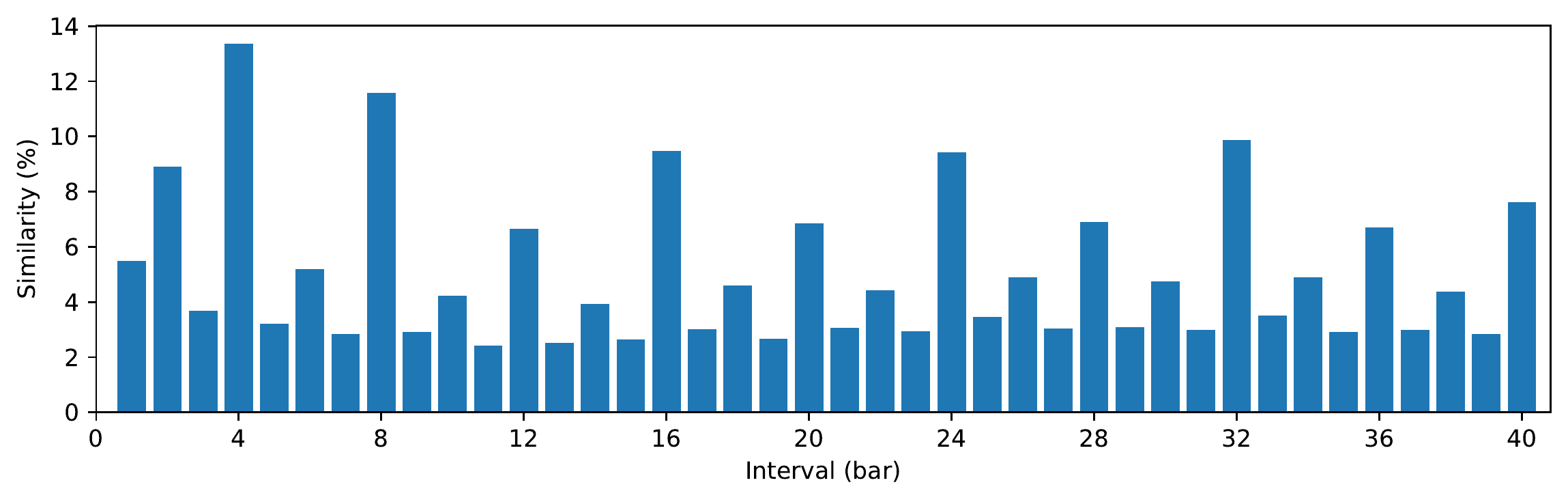}
		% \fbox{\rule[-.5cm]{0cm}{4cm} \rule[-.5cm]{14cm}{0cm}}
		\caption{The similarity distribution of the training set used in our experiments. It is calculated over the melody track because melody best shows the structure of a song. We omit the similarity when $t=0$ because the similarity between a bar and itself is always $1.0$.}
		\label{fig:similarity_statistics}
	\end{figure}

	In this paper, we call the distribution of the similarity with respect to the bar intervals the similarity distribution, and show that of the training data in \autoref{fig:similarity_statistics}. 
	As we can see, it shows an obvious periodical pattern -- a music bar tends to be more similar to its previous $2$ bars, and also to the previous $4$-th bar or its multipliers in most cases.
	We further conduct the similarity statistics on different datasets involving music of various genres and styles. The results shown in \autoref{sec:app_similarity} interestingly indicate that this pattern is universally applicable to the music of the great diversity. We believe that it can be regarded as a general rule applicable to most music in our daily life.
	
	Based on the statistical results, we carefully select 8 bars -- the previous 1st, 2nd, 4th, 8th, 12th, 16th, 24th, and 32nd bars, as the default structured-related bars, since they can cover the most similar bars in most cases. The number of selected structure-related bars is a trade-off between efficiency and information richness. One can select more or other bars according to the computation resources and the specific similarity distribution of the used dataset. We have to admit that the selected bars cannot always cater to any song, but as long as it covers most cases, it can already enable Museformer to generate music with better structures.
	
	% Considering that music in most cases is periodical sequence -- it usually repeats or imitates the previous pieces at certain intervals (see the appendix for a simple example), we propose to conduct a similarity statistics over the human-made music, to see the distribution of similarity with respect to the bar intervals. \autoref{fig:similarity_statistics} shows the result conducted on Lakh MIDI, a large-scale, widely-used and multi-genre symbolic music dataset.
	
% 	Please note that the selected 8 bars are only one case that fits for most situations, and we can also select other bars as the structure-related bars according to the statistics on a specific dataset.
	
% 	\begin{figure}
% 		\centering
% 		\includegraphics[width=0.9\textwidth]{fig/repetition_stat2}
% 		% \fbox{\rule[-.5cm]{0cm}{4cm} \rule[-.5cm]{14cm}{0cm}}
% 		\caption{Statistical results on the repetition intervals.}
% 		\label{fig:repetition_statistics}
% 	\end{figure}

	\subsection{Merits of Museformer}
	\label{subsec:analysis}
	
	We explain why Museformer is suitable for the music generation task.
	
	First, the receptive fields of the fine-grained attention comply with the characteristics of music and can cover most of the structure-related information.
	Unlike previous models that adopt fixed token-level patterns \cite{beltagy2020longformer,chen2021tfmls} or learn content-based patterns \cite{Kitaev2020Reformer, wu2021smart}, which may not cover the most important ranges to be referred to for generating music structures, we directly use the structure-related bars derived from human-made music. These bars can include both neighboring ones and distant ones, which enables Museformer to generate both short-term and long-term music structures.
	
	% First, the structure-related bar selection ensures that FC-Attention can capture both the short-term and the long-term structures of music. Compared to most long-sequence Transformers that only focus on the most recent content (e.g., recurrent Transformers and sliding window attention), the selected bars include both the neighboring ones to capture the short-term structures and the distant ones to capture the long-term structures.
	
	Second, the coarse-grained attention can preserve necessary information for generating better music. 
	In general, FC-Attention can be categorized into sparse attention, but unlike conventional sparse attention that simply drops a large amount of information, which may limit the model's capability, especially for music where mutual connections abound, the coarse-grained attention preserves the information of the other bars so as to provide rich clues for generation.
	
	% Secondly, the combination of the fine-grained attention and the coarse-grained attention makes it capable of concentrating on the structure-related bars, and also preserving the necessary information of the other bars through the summarization. In contrast, sparse attention models leverage strong inductive biases to make each specific token only attend to a subset of tokens and simply drop the others, which may loses necessary information of the whole sequence.
	
	Third, the combination of the fine- and coarse-grained attention enables Museformer to handle long music sequences efficiently. Compared to the number of all the music tokens, the number of tokens in the structure-related bars and the summary tokens is much smaller, so the memory consumption and the running time are greatly reduced. It is essential for this task where sequences are quite long.

	\section{Experiments}
	\label{sec:exp}
	
	\subsection{Experiment Settings}
	\label{subsec:exp_settings}
	We give a brief introduction to the experiment settings. Please refer to \autoref{sec:app_exp_settings} for more details.
	
	\paragraph{Dataset and Music Representation}
	We conduct our experiments on the widely used Lakh MIDI (LMD) dataset \cite{raffel2016learning}, which contains multi-instrument music in the format of MIDI. After preprocessing, the final dataset contains \num{29940} songs (\num{1727} hours), each song contains $95$ bars on average. Following \cite{ren2020popmag}, various instruments are merged into the $6$ basic ones namely square synthesizer, piano, guitar, string, bass, and drum, with the square synthesizer playing the melody. We use a REMI-like representation method \cite{huang2020pop} to transfer MIDIs into token sequences, where the musical information (instrument, bar line, note position, pitch, duration, etc.) is represented in separate tokens (see \autoref{fig:score} for an example). Each song is represented by \num{15042} tokens on average, and each bar is represented by $158$ tokens on average. We randomly split all the songs by $8/1/1$ for training/validation/test, respectively.
	
	\paragraph{Implementation}
	Museformer is implemented with PyTorch\footnote{\url{https://pytorch.org}} and fairseq\footnote{\url{https://github.com/pytorch/fairseq}}. For the efficient computation of FC-Attention, we write CUDA kernels to construct the attention layouts for each sample, and then transfer the layouts into a blocksparse form and compute with SparTA\footnote{\url{https://github.com/microsoft/SparTA}} \cite{SparTA2022}.
	
	\paragraph{Model and Training Configurations}
	The core parameters of Museformer include layer number $= 4$, hidden size $= 512$, number of attention heads $= 8$, and FFN hidden size $=$ \num{2048}.
	During training, the batch size is set to $4$ songs. Following \cite{vaswani2017attention}, we use the Adam optimizer with $\beta_1=0.9$, $\beta_2=0.98$ and $\varepsilon=10^{-9}$, and the learning rate is warmed up over the first \num{16000} steps linearly to a peak value of $5\times 10^{-4}$, and then decreases based on the inverse square root of the steps. The L2 weight decay is set to $0.01$. During inference, we use top-$k$ sampling with $k=8$. Generation continues until the end-of-sentence token is generated or it reaches the maximum length of \num{20480}.

	\paragraph{Compared Models}
	We compare Museformer with $4$ representative Transformer-based models, most of which have been adopted in music generation: 
	\begin{itemize}[leftmargin=*]
	\item \textbf{Music Transformer} \cite{huang2018music}: a vanilla Transformer with a memory-efficient ``skewing'' relative position embedding implementation. 
	\item \textbf{Transformer-XL} \cite{dai2019xl}: a recurrent Transformer that encodes the sequence chunk by chunk and uses the gradient-stopped representations of previous chunks as the memory. 
	\item \textbf{Longformer} \cite{beltagy2020longformer}: a model with a sliding window sparse attention. 
	\item \textbf{Linear Transformer} \cite{katharopoulos2020transformers}: a model that uses a kernel-based attention of linear complexity. 
	\end{itemize}
	All the compared models are set with comparable hyper-parameters as Museformer. For Music Transformer that uses the full attention and thus cannot model long sequences at once due to memory limit, we chunk each song into multiple samples during training, and apply the model to generate long sequences during validation and inference to test its generalization on long music sequences.

	\paragraph{Objective Evaluation}
	We use the following objective metrics to evaluate the models:
	\begin{itemize}[leftmargin=*]
	    \item \textbf{Perplexity (PPL)}: a common metric to measure whether a generative model can correctly predict future tokens. The smaller, the better. To see the models' performances on different lengths, it is calculated on the first \num{1024}, \num{5120}, or \num{10240} tokens of each sample.
	    \item \textbf{Similarity Error (SE)}: the error between the similarity distribution of training data and generated music, to evaluate the models' ability to generate music with realistic structures. It is defined as
	   % \begin{equation}
	   %     e = \sum_{d=1}^{D} \bigg | \frac{1}{| N_{\text{gen}} |} \sum_{i=1}^{| N_{\text{gen}} |} \hat{s}_{i,d} -  
	   %     \frac{1}{|N_{\text{train}}|} \sum_{i=1}^{|N_{\text{train}}|} s_{j,d} \bigg |
	   % \end{equation}
	   \begin{equation}
	        \text{SE} = \frac{1}{T} \sum_{t=1}^{T} | \hat{L}_{t} - L_{t} | ,
	    \end{equation}
	    where $\hat{L}_{t}$ and $L_{t}$ are the average similarities (defined in \autoref{eq:similarity_distribution}) of the generated music and the training data respectively. We set $T=40$ in our experiments, and $\hat{L}_{t}$ is calculated on $100$ generated music pieces for each model. The smaller the value is, the more similar the structures of the generated music are to human-made music. 
	    % We also calculate a \textbf{normalized similarity error (NSE)}, where the two similarities are L1-normalized over the $D$ distances.
	\end{itemize}
	% We use \textbf{Perplexity (PPL)} to evaluate the performance, which is a very common metric in the generation tasks to measure whether a model can correctly predict the tokens. The lower the value is, the better the model can correctly predict.
	% In our experiment, perplexity is calculated on the test set and over the first \num{1024}, \num{5120}, or \num{10240} tokens of each sample.
	
	% 2) \textbf{Beat STD}, \textbf{downbeat STD} and \textbf{downbeat salience}. These are the metrics adopted by \citet{huang2020pop} to show the rhythmic features. To calculate these metrics, we first use MuseScore to render the generated samples into audio clips, and apply the joint beat and downbeat tracking model \cite{bock2016downbeat} to identify beats and downbeats (including their time point and the salience probability). Beat STD and downbeat STD are the standard deviations of the beat and downbeat intervals respectively, and which evaluate the flexibility of the rhythms. Downbeat salience is the mean value of the downbeat salience probability, and it evaluates the salience of the rhythm.
	
	\paragraph{Subjective Evaluation}
	The most canonical way to evaluate a music generation model is the human listening test. We apply each model to randomly generate $100$ music pieces, and invite $10$ people, where $7$ of them have music-related learning experiences, to score these music pieces. Specifically, for each participant, we randomly construct $5$ groups, where each group contains $5$ music pieces that are generated by Museformer and the $4$ compared models, respectively. Participants are asked to score these music pieces from $1$ (lowest) to $10$ (highest) over the following subjective metrics: 
	\begin{itemize}[leftmargin=*]
	    \item \textbf{Musicality}: whether it is pleasant and interesting, and real enough just like human-made music. 
	    \item \textbf{Short-term structure}: whether it shows good structures in neighboring content, such as good repetitions and reasonable development. 
	    \item \textbf{Long-term structure}: whether it shows good structures in long distances, such as song-level repetitions and long-distance connections. 
	    \item \textbf{Overall}: an overall score. We also calculate \textbf{preference score} based on this overall score, which is defined as the ratio of winning times (obtaining the highest overall score within a group).
	\end{itemize}
	
	\subsection{Comparison with Previous Models}
	\autoref{tab:objective_result} shows the results of the objective evaluation. 
	As we can see: 
	1) Music Transformer achieves a comparable PPL as other models on short music sequences (\num{1024} tokens) but undergoes a severe deterioration on longer sequences, which implies that a model trained on short music sequences cannot well generalize to long music sequences. Accordingly, an appropriate long-sequence Transformer model is needed for modeling the long music sequences. 
	2) Although the receptive field of Linear Transformer can cover the whole sequence, its PPLs show no superiority compared to other models, which may be because the kernel-based attention cannot accurately capture the complex correlations of music. 
	3) The proposed Museformer can consistently achieve the best PPLs on various sequence lengths, especially on a larger length, which demonstrates the effectiveness of Museformer on the music generation task. 
	4) The results on SE demonstrate that the music generated by Museformer has structures that are most similar to the human-made music. We provide the similarity distribution for each model and more discussion in \autoref{sec:generated_similarity_analysis}.
	
	\begin{table}
	    \caption{The results of objective evaluation and ablation study. The numbers in the parentheses for PPLs are sequence lengths.}
		\label{tab:objective_result}
		\begin{center}
			\begin{tabular}{lcccc}
				\toprule
				& PPL (\num{1024}) & PPL (\num{5120}) & PPL (\num{10240}) & SE (\%) \\
				\midrule
				Music Transformer & $1.66$ & $1.77$ & $2.55$ & $2.49$  \\
				Transformer-XL & \bm{$1.64$} & $1.45$ &  $1.43$ & $15.66$ \\
				Longformer & $1.65$ & $1.46$ & $1.45$ & $5.25$ \\
				Linear Transformer & $1.86$ & $1.67$ & $1.64$ & $1.97$  \\
				\midrule
				Museformer (ours) & \bm{$1.64$} & \bm{$1.41$} & \bm{$1.35$} & \bm{$0.95$} \\
				\quad w/o coarse-grained  & $1.65$ &  $1.42$  &  $1.38$  & $1.08$ \\
				\quad w/o bar selection   & $1.65$  & $1.43$  &  $1.39$  & $6.39$ \\
				\bottomrule
			\end{tabular}
		\end{center}
	\end{table}
	
	Furthermore, we present the results of the subjective evaluation in \autoref{tab:subjective_results_score}, which show that Museformer achieves the best performance. Specifically, 1) Museformer gets the highest scores on all the metrics. 2) On the structure-related metrics, especially the long-term structure, Museformer exceeds other models by a large gap, indicating that the proposed FC-Attention can empower the model to capture the correlations in distant bars.
	
	\begin{table}
	    \caption{The results of subjective evaluation. ST and LT stand for short-term and long-term respectively. For all the subjective metrics, mean $\pm$ standard deviation is reported. Pref stands for preference score.}
		\label{tab:subjective_results_score}
		\begin{center}
			\begin{tabular}{lccccc}
				\toprule
				    & Musicality & ST structure & LT structure & Overall & Pref \\
				\midrule
				Music Transformer & $6.00 \pm 2.21$  & $6.90 \pm 1.76$  &  $5.30 \pm 2.58$  &  $5.90 \pm 1.90$  &  $0.20$  \\
				Transformer-XL & $6.10 \pm 2.19$ & $7.40 \pm 1.81$  & $6.26 \pm 2.78$ & $6.44 \pm 2.01$  &  $0.34$ \\
				Longformer & $6.46 \pm 1.81$ & $7.60 \pm 1.47$ & $6.18 \pm 2.54$ & $6.44 \pm 1.72$  &  $0.24$ \\
				Linear Transformer  & $6.06 \pm 1.99$ & $6.92 \pm 2.03$ & $5.78 \pm 2.64$ & $6.30 \pm 1.84$  &  $0.24$ \\
				\midrule
				Museformer (ours)  & $\mathbf{6.88} \pm \mathbf{1.95}$ & $\mathbf{7.86} \pm \mathbf{1.51}$ & $\mathbf{6.72} \pm \mathbf{2.74}$ & $\mathbf{7.12} \pm \mathbf{1.81}$  &  $\mathbf{0.46}$ \\
				\bottomrule
			\end{tabular}
		\end{center}
	\end{table}
	
	We further do pairwise comparisons over the subjective evaluation results. \autoref{tab:significance} shows the number of wins/ties/losses based on the overall scores, as well as the $p$-values of the Wilcoxon signed rank test. Museformer obtains more wins than the compared models, and the $p$-values indicate that Museformer achieves statistically significant improvements ($p < 0.05$).
	
	\begin{table}
	    \caption{The results of pairwise comparisons based on the subjective overall scores.}
		\label{tab:significance}
		\begin{center}
			\begin{tabular}{lcccc}
				\toprule
				& Wins & Ties & Losses & $p$-value \\
				\midrule
				Museformer VS Music Transformer & $33$ & $7$ & $10$ & $0.0003$ \\
				Museformer VS Transformer-XL  & $25$ &  $14$  &  $11$ &  $0.0375$\\
				Museformer VS Longformer & $28$  &  $8$  & $14$  & $0.0424$ \\
				Museformer VS Linear Transformer & $29$ & $7$ & $14$ & $0.0128$ \\
				\bottomrule
			\end{tabular}
		\end{center}
	\end{table}

	\subsection{Ablation Study}
	
	We verify the effectiveness of the Museformer components as follows: 1) To see the effectiveness of the coarse-grained attention, we compare Museformer with the setting \textbf{w/o coarse-grained}, where each music token attends to the music tokens in the structure-related bars and its previous ones in the current bars, and no summary token of any bar. 2) To see the effectiveness of the structure-related bar selection for the fine-grained attention, we compare Museformer with the setting \textbf{w/o bar selection}, where we use the most recent $8$ bars for the fine-grained attention instead of the structure-related bars selected via the similarity statistics.
	
	From the results presented in the bottom block of \autoref{tab:objective_result}, we can observe that: 
	1) Museformer consistently achieves better performances on both PPLs and SE compared to the two ablation settings, which demonstrates the effectiveness of the coarse-grained attention and the structure-related bar selection. 
	2) The coarse-grained attention preserves the necessary information of the bars other than the structured-related bars and can consistently help decrease the PPLs. 
	3) The contribution of the structure-related bar selection increases as the sequences get longer. This is reasonable because that in a longer music sequence, there tends to be more bars and more long-term structures, and the fine-grained attention can directly capture the correlations in the distant structure-related bars, which helps make more accurate predictions. 
	4) Embodied on SE, the bar selection is contributory to the structures of generated music. Please refer to \autoref{sec:generated_similarity_analysis} for more details. 
	In addition to the above two settings, we have also tried disabling the fine-grained attention for previous bars, where a music token only directly attends to its previous tokens within its bar and attends to all the previous bars via the summary tokens. The result indicates that only summarized information and no precise token-level information for the previous bars is insufficient for the model to generate coherent music.
	
	\subsection{Complexity Analysis}
	
	Suppose the sequence length is $n$, the average bar number is $b$, the average bar length is $m$, and the number of selected structure-related bars is $k$. For FC-Attention, the time complexity of the summarization step is $O(n)$, and that of the aggregation step is $O\big( (km + b)n \big)$, so the overall complexity is $O\big((km+b)n\big)=O\big((km+n/m)n\big)$. Although the complexity is still proportional to the square of $n$, the typically large divider $m$ (mostly exceeds \num{100}) greatly reduces the complexity. Thus, the specific complexity is between linear and quadratic. Note that the computation of FC-Attention depends on the input content, i.e., the number of bars and the number of tokens in each individual bar, the complexity cannot be precisely formulated, and the efficiency in real applications is more noteworthy.
	
\begin{figure}
        \begin{subfigure}[b]{0.49\textwidth}
            \centering
            \includegraphics[width=\textwidth]{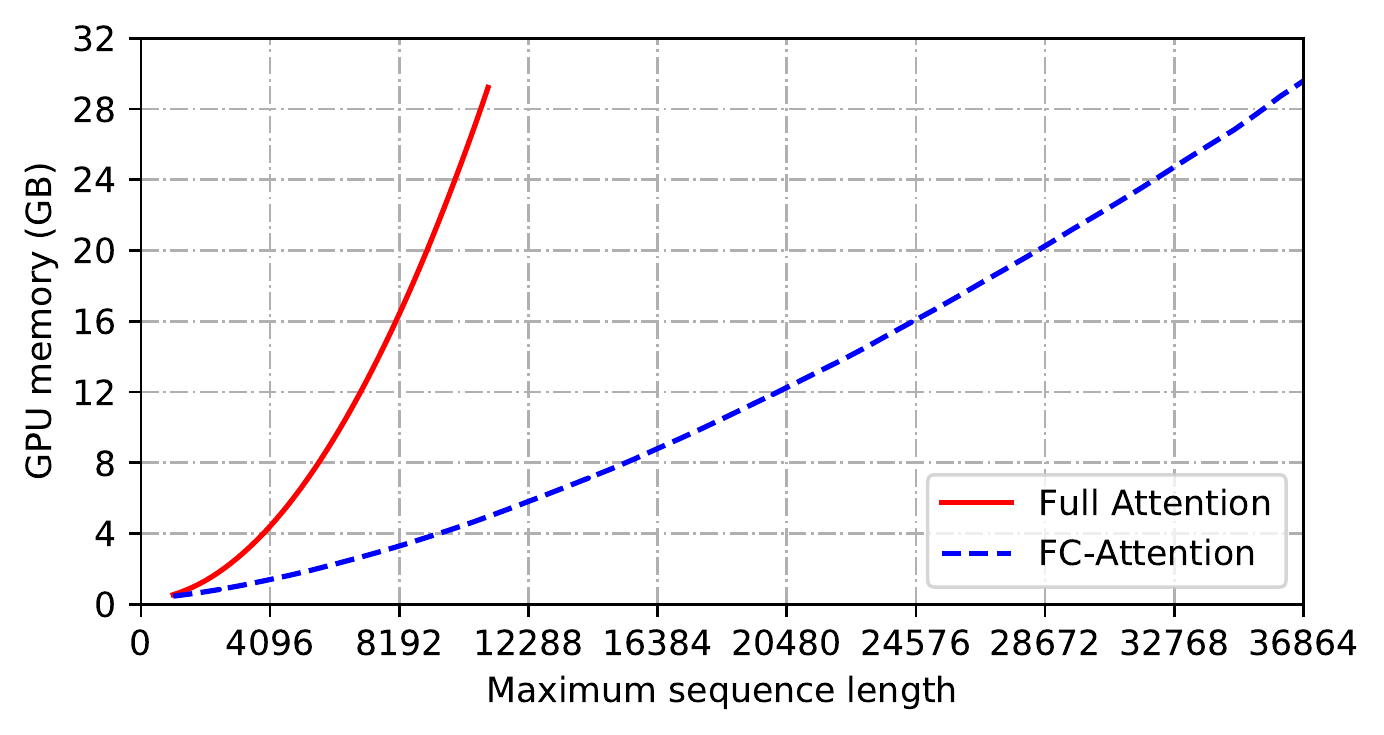}
            \caption{GPU memory consumption.}
            \label{fig:eff1}
        \end{subfigure}
        \hfill
        \begin{subfigure}[b]{0.49\textwidth}
            \centering
            \includegraphics[width=\textwidth]{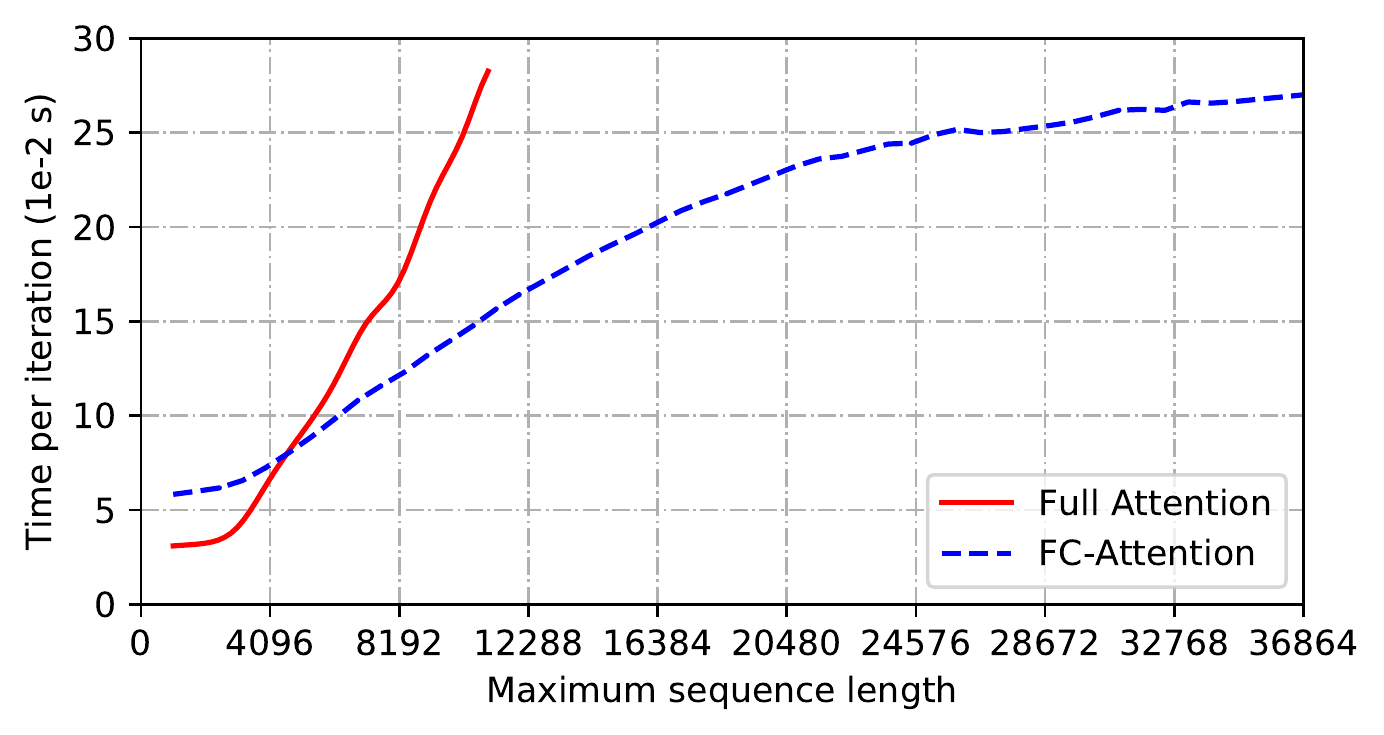}
            \caption{Running time per iteration.}
            \label{fig:eff2}
        \end{subfigure}
		\caption{Memory consumption and running time of Museformer with FC-Attention compared to its full attention counterpart.}
		\label{fig:memory_consumption}
	\end{figure}

	To see the efficiency of FC-Attention in real applications, we train Museformer and its full attention counterpart on the validation set, and record their memory consumption and running time. With batch size $= 1$, we increase the maximum sequence length until we use up the $32$GB memory of an Nvidia V100 GPU, and record the peak memory consumption\footnote{Obtained by torch.cuda.max\_memory\_allocated().} and the running time for one-epoch training. \autoref{fig:memory_consumption} shows the results. We observe that the memory consumption of Museformer increases at a nearly linear rate with respect to the maximum sequence length, and can process over $3\times$ longer music sequences than its full attention counterpart, making it capable of generating full-song music. Museformer also achieves faster training time when the maximum sequence length is greater than around \num{5000}, which is common for music sequences.

	\subsection{Case Study}
	
\begin{figure}[htbp]
		\centering
		\includegraphics[width=0.85\textwidth]{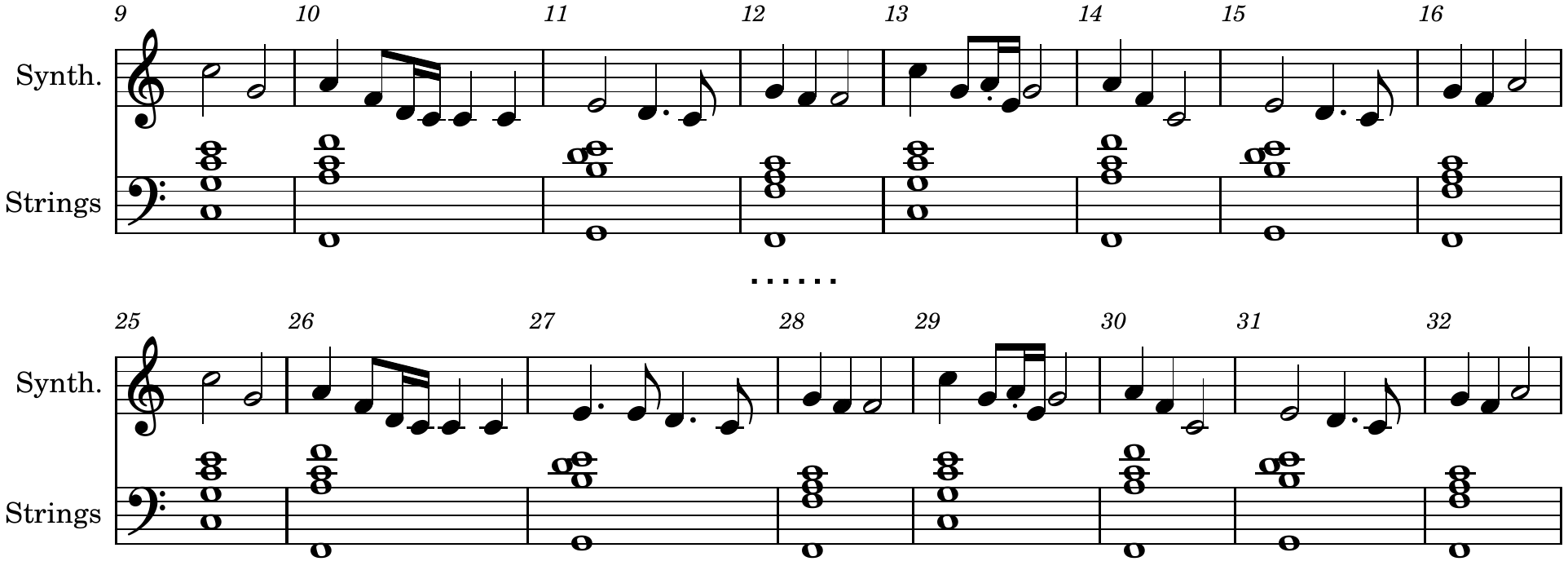}
		\caption{A snippet of a generated song.}
		\label{fig:case_study}
	\end{figure}

	\autoref{fig:case_study} shows a snippet of a song generated by Museformer.
	As we can see, on the \textit{Strings} track, the 13th - 16th bars repeat the 9th - 12th bars with an interval of 4 bars, which is a short-term structure. The 25th - 32nd bars repeat the 9th - 16th bars with an interval of 16 bars, and there are reasonable variations in the 27th bar compared to the 11th bar, which shows a long-term structure. This case demonstrates that Museformer can generate music with both short-term and long-term structures, and not only exact repetitions but also some variations. In addition to the two exhibited tracks, other tracks, such as piano and drum, have more variations on the two similar segments, which creates reasonable development of music. Please refer to our demo page for more information.
	
	\section{Conclusion}
	
	To solve the long sequence modeling and music structure modeling challenges in symbolic music generation, we propose Museformer with a novel fine- and coarse-grained attention. 
	The fine-grained attention is applied to the structure-related bars for learning the structure-related correlations, and the coarse-grained attention is applied to the summarization of the other bars for getting a sketch of them.
	We propose to select the structure-related bars via bar-pair similarity statistics.
	Experimental results show that Museformer is efficient and can generate music with good quality and structures. 
	
	Museformer is not perfect yet, and we would like to discuss about its limitations and possible future explorations. 
	First, since Museformer takes random samplings during inference and does not receive manual control, it can hardly ensure that every generated music piece is well-structured in an expected way. Techniques to enhance its reliability and controllability can be further explored. 
	Furthermore, the musicality and creativity of the generated music are still far behind that of human-made music, which remains a problem for all the existing music generation models. We believe that more sophisticated music representation and large models trained on large-scale data can help alleviate this problem.
	Finally, we anticipate Museformer's adaptation to more tasks and domains. It can be easily and reasonably applied to music understanding tasks. For NLP tasks, since natural languages do not have the periodical patterns as music, how to determine the semantically related sentences, paragraphs, or documents is an interesting challenge to solve.

	\section*{Acknowledgments}
	
	This work is supported by National Natural Science Foundation of China (No. 61872172). We thank the scorers in the subjective evaluation, the SparTA group at Microsoft Research Asia, and many other people from the Websoft group at Nanjing University and Microsoft for their kind help.

{
		% \small
		\bibliographystyle{IEEEtranN}
		\bibliography{ref}
	}

	\newpage
	\appendix

\section{Similarity Statistics on Different Datasets}
\label{sec:app_similarity}

\begin{figure}[htbp]
		\centering
		\begin{subfigure}[b]{\textwidth}
            \centering
            \includegraphics[width=\textwidth]{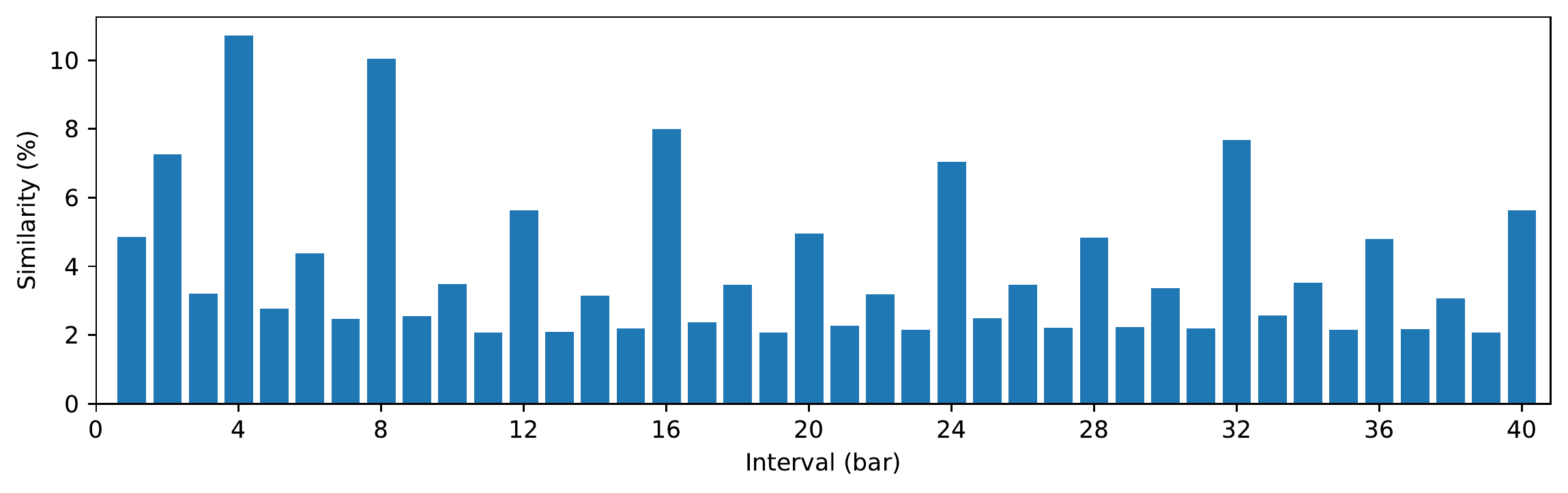}
            \caption{Melody track.}
            \label{fig:ds_lmdn_all_melody}
        \end{subfigure}
		
        \begin{subfigure}[b]{0.49\textwidth}
            \centering
            \includegraphics[width=\textwidth]{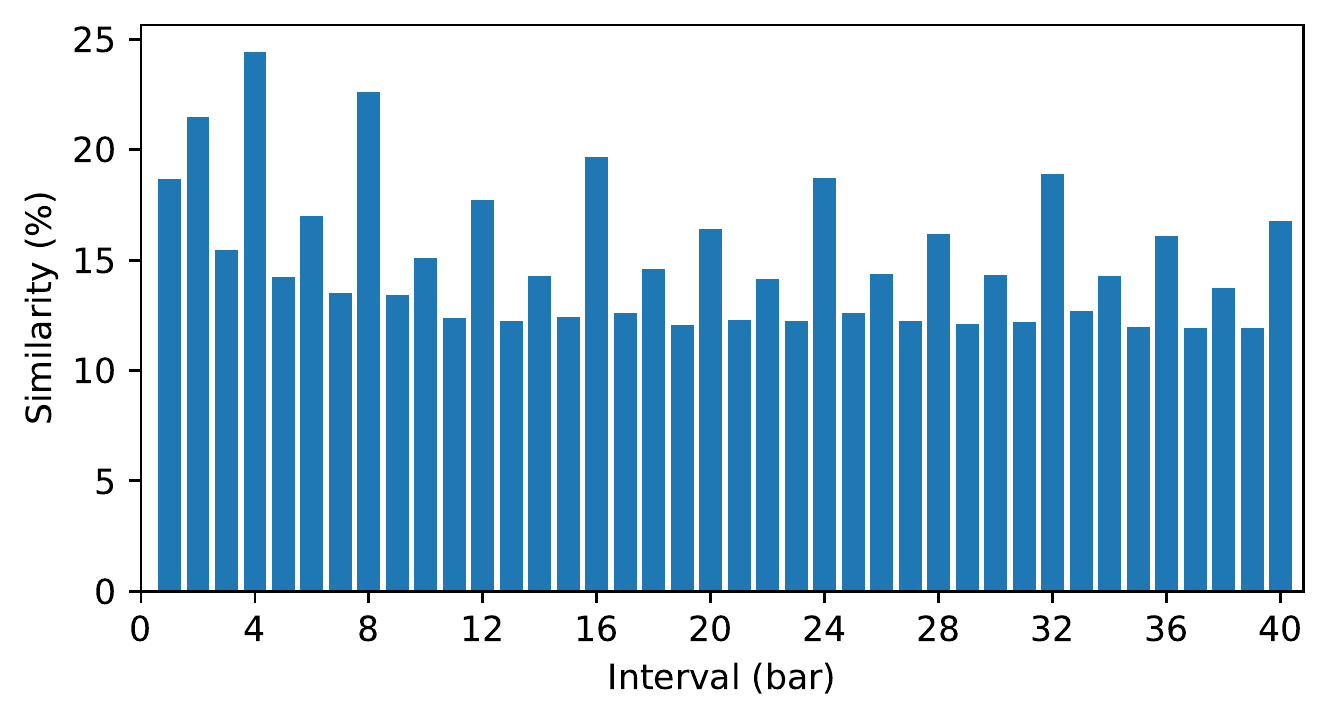}
            \caption{All tracks.}
            \label{fig:ds_lmdn_all_all}
        \end{subfigure}
        \hfill
        \begin{subfigure}[b]{0.49\textwidth}
            \centering
            \includegraphics[width=\textwidth]{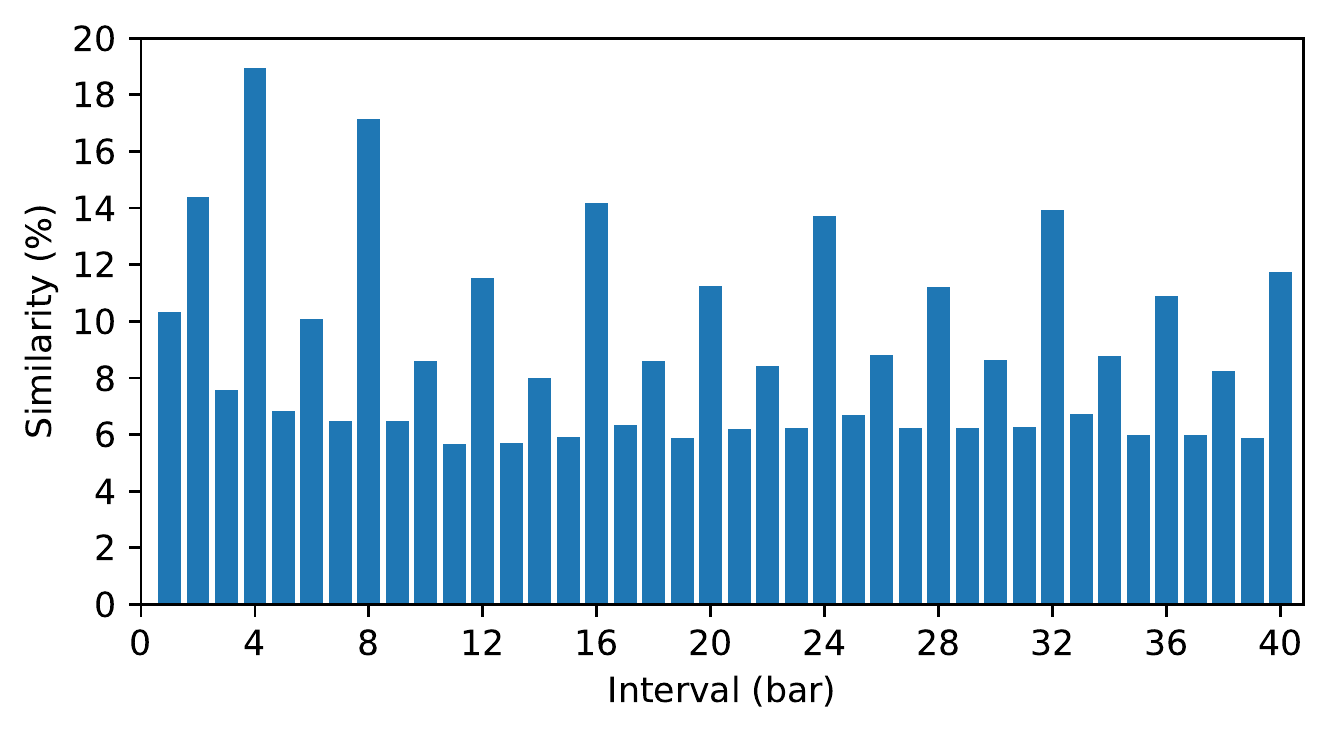}
            % \fbox{\rule[-.5cm]{0cm}{4cm} \rule[-.5cm]{4cm}{0cm}}
            \caption{Piano track.}
            \label{fig:ds_lmdn_all_piano}
        \end{subfigure}
        
        \begin{subfigure}[b]{0.49\textwidth}
            \centering
            \includegraphics[width=\textwidth]{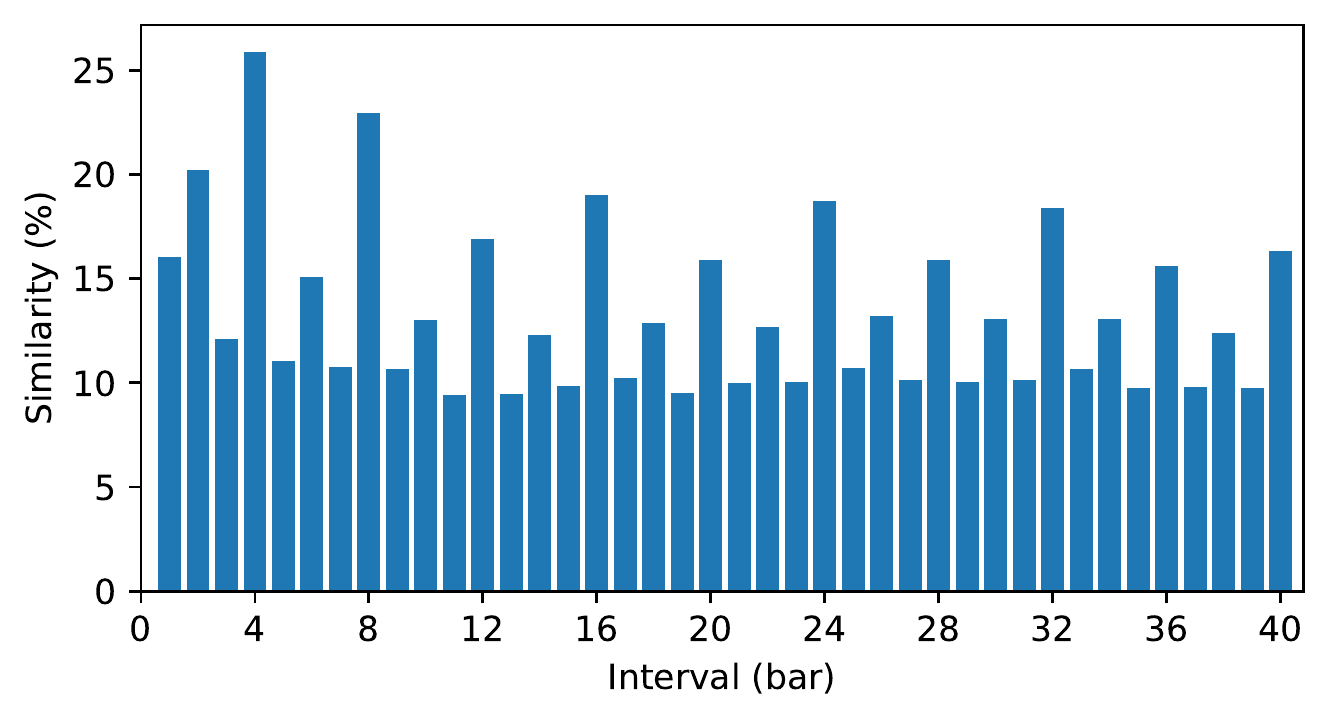}
            % \fbox{\rule[-.5cm]{0cm}{4cm} \rule[-.5cm]{4cm}{0cm}}
            \caption{Guitar track.}
            \label{fig:ds_lmdn_all_guitar}
        \end{subfigure}
        \hfill
        \begin{subfigure}[b]{0.49\textwidth}
            \centering
            \includegraphics[width=\textwidth]{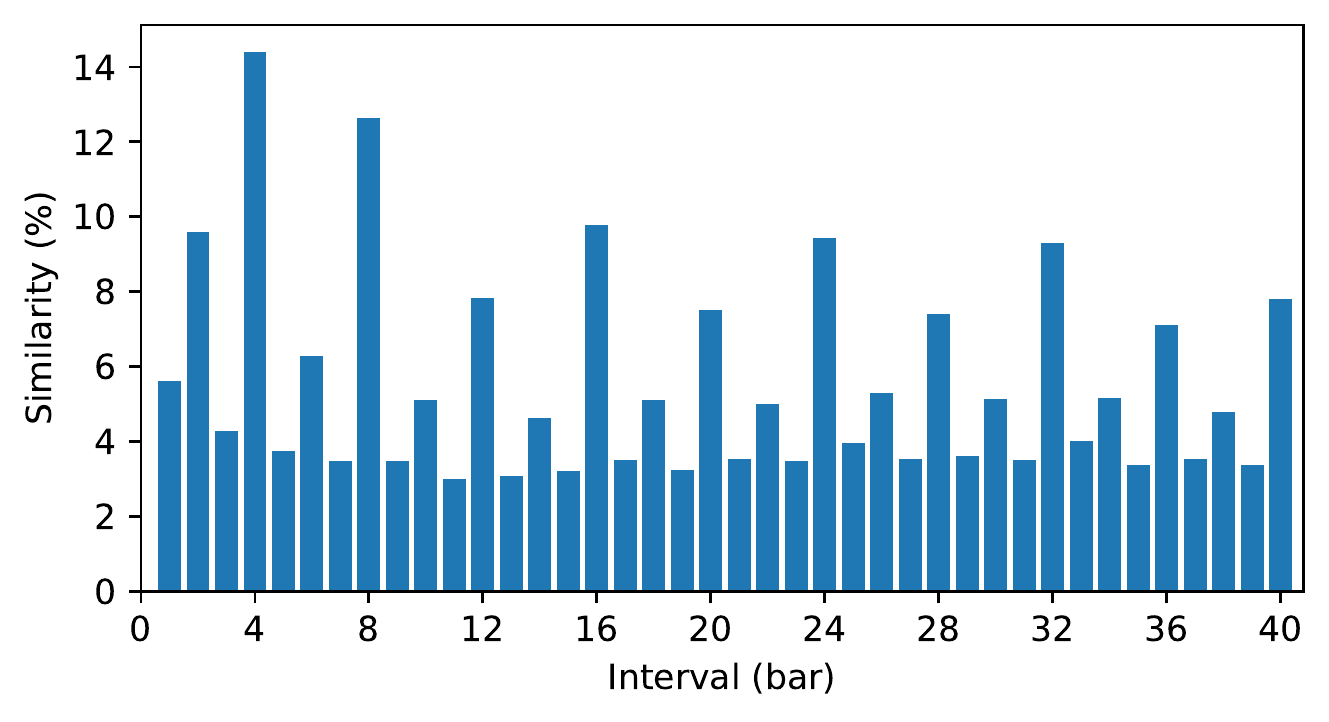}
            % \fbox{\rule[-.5cm]{0cm}{4cm} \rule[-.5cm]{4cm}{0cm}}
            \caption{String track.}
            \label{fig:ds_lmdn_all_string}
        \end{subfigure}
        
        \begin{subfigure}[b]{0.49\textwidth}
            \centering
            \includegraphics[width=\textwidth]{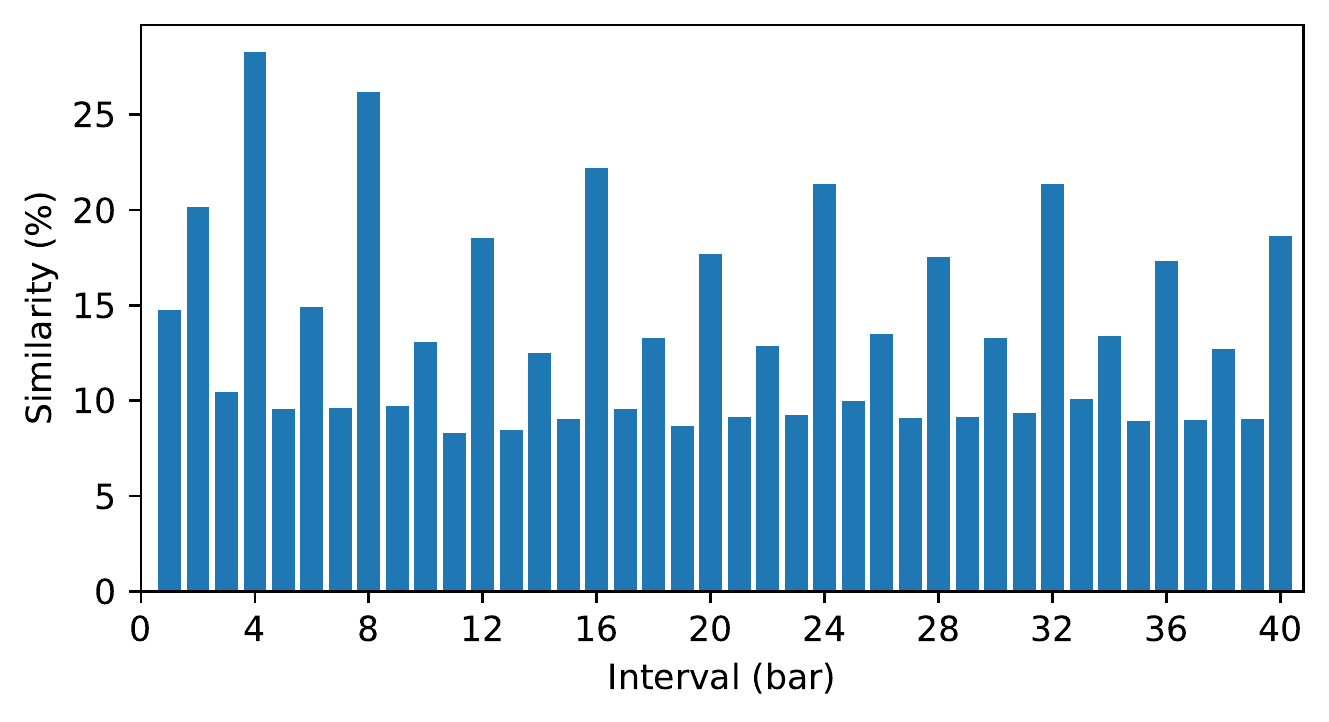}
            % \fbox{\rule[-.5cm]{0cm}{4cm} \rule[-.5cm]{4cm}{0cm}}
            \caption{Bass track.}
            \label{fig:ds_lmdn_all_bass}
        \end{subfigure}
        \hfill
        \begin{subfigure}[b]{0.49\textwidth}
            \centering
            \includegraphics[width=\textwidth]{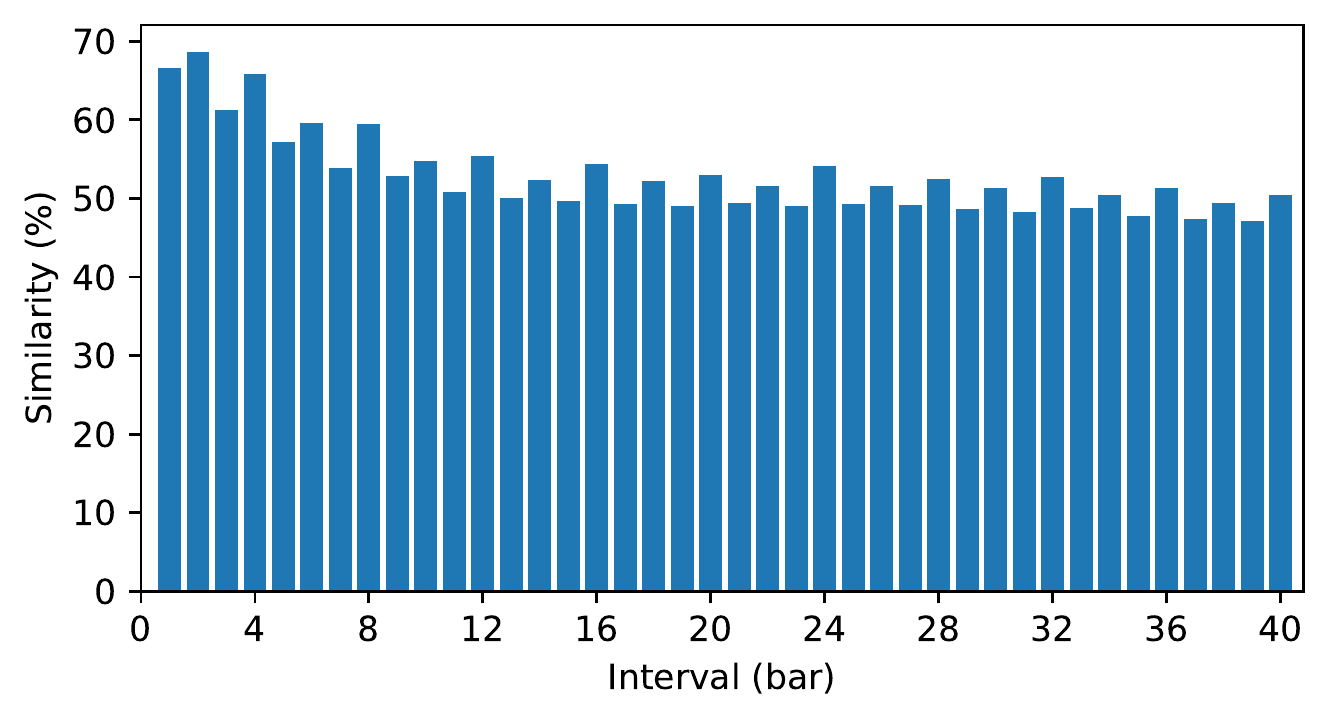}
            % \fbox{\rule[-.5cm]{0cm}{4cm} \rule[-.5cm]{4cm}{0cm}}
            \caption{Drum track.}
            \label{fig:ds_lmdn_all_drum}
        \end{subfigure}
		\caption{The similarity distribution of the LMD dataset we use.}
		\label{fig:ds_lmdn}
	\end{figure}

    We conduct the similarity statistics introduced in \autoref{subsec:bar_selection} on the whole LMD dataset that we use, and exhibit the results in \autoref{fig:ds_lmdn}. We can observe that the periodical structure pattern, that a music bar tends to be more similar to its previous $2$ bars, and also to the previous $4$-th bar or its multipliers in most cases, is also satisfied on this set of music. Specifically, this rule holds on all of the instrument tracks except the drum track. This makes sense because the percussive instruments usually play the same rhythmic patterns over and over throughout a song.

\begin{figure}[htbp]
		\centering
		\begin{subfigure}[b]{0.32\textwidth}
            \centering
            \includegraphics[width=\textwidth]{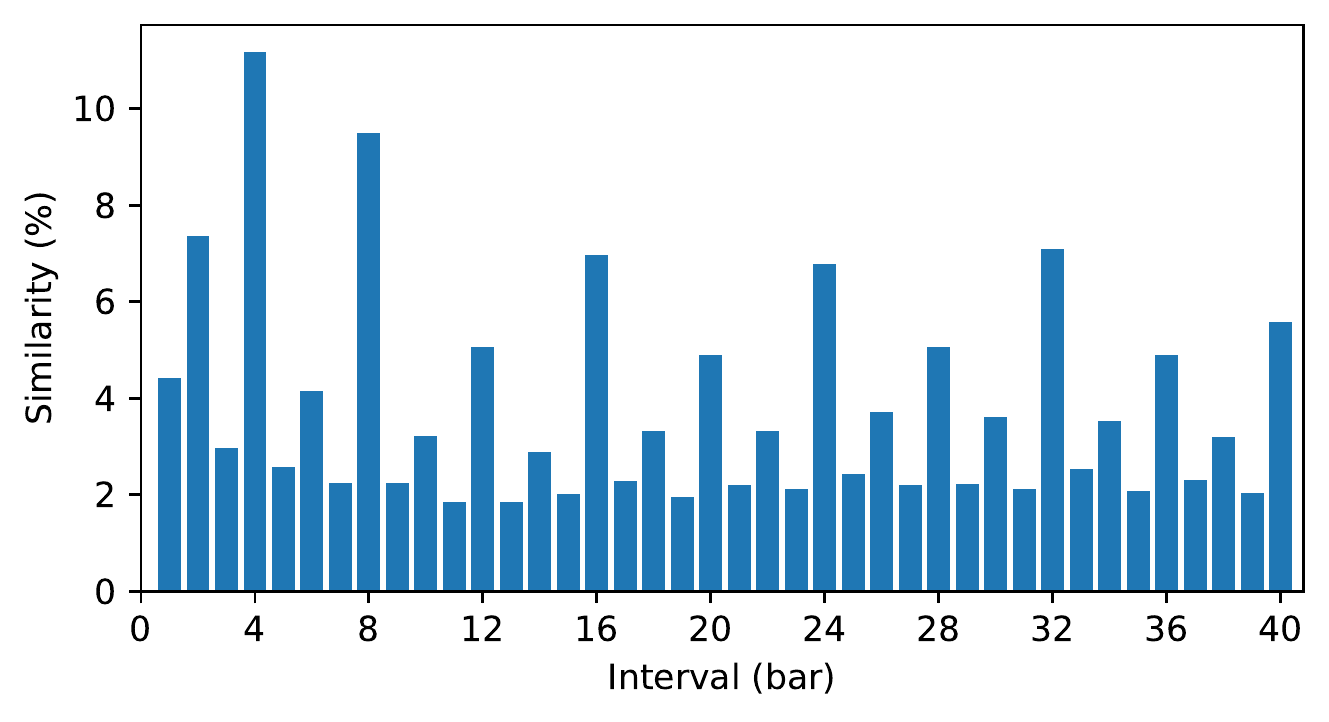}
            \caption{All genres.}
        \end{subfigure}
		\hfill
        \begin{subfigure}[b]{0.32\textwidth}
            \centering
            \includegraphics[width=\textwidth]{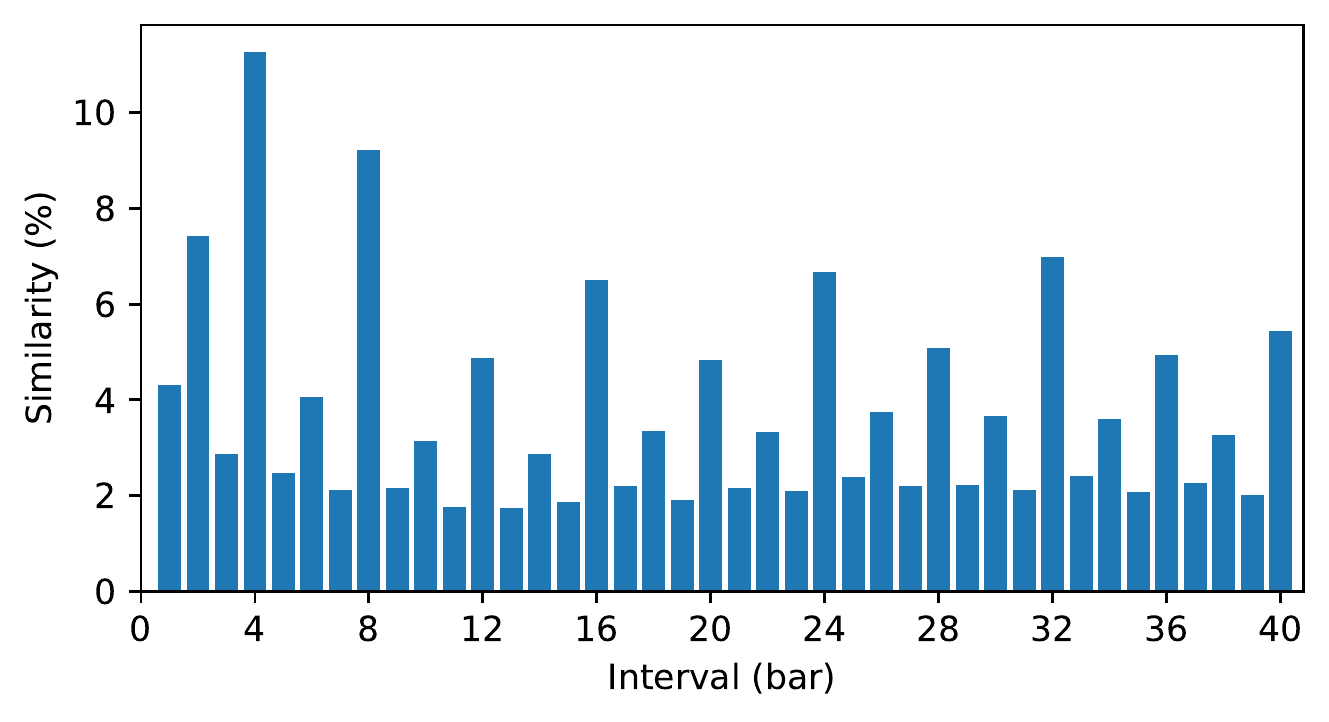}
            \caption{Pop/rock.}
        \end{subfigure}
        \hfill
        \begin{subfigure}[b]{0.32\textwidth}
            \centering
            \includegraphics[width=\textwidth]{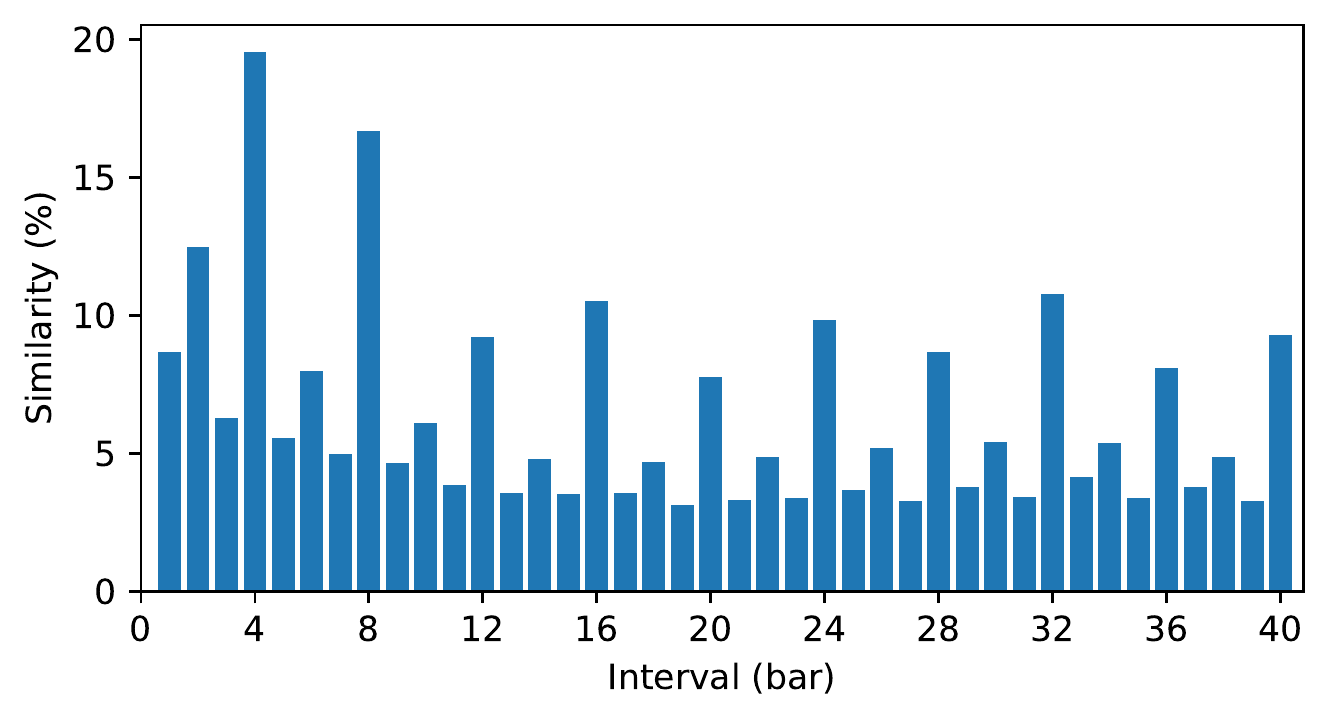}
            % \fbox{\rule[-.5cm]{0cm}{4cm} \rule[-.5cm]{4cm}{0cm}}
            \caption{Electronic.}
        \end{subfigure}
        
        \begin{subfigure}[b]{0.32\textwidth}
            \centering
            \includegraphics[width=\textwidth]{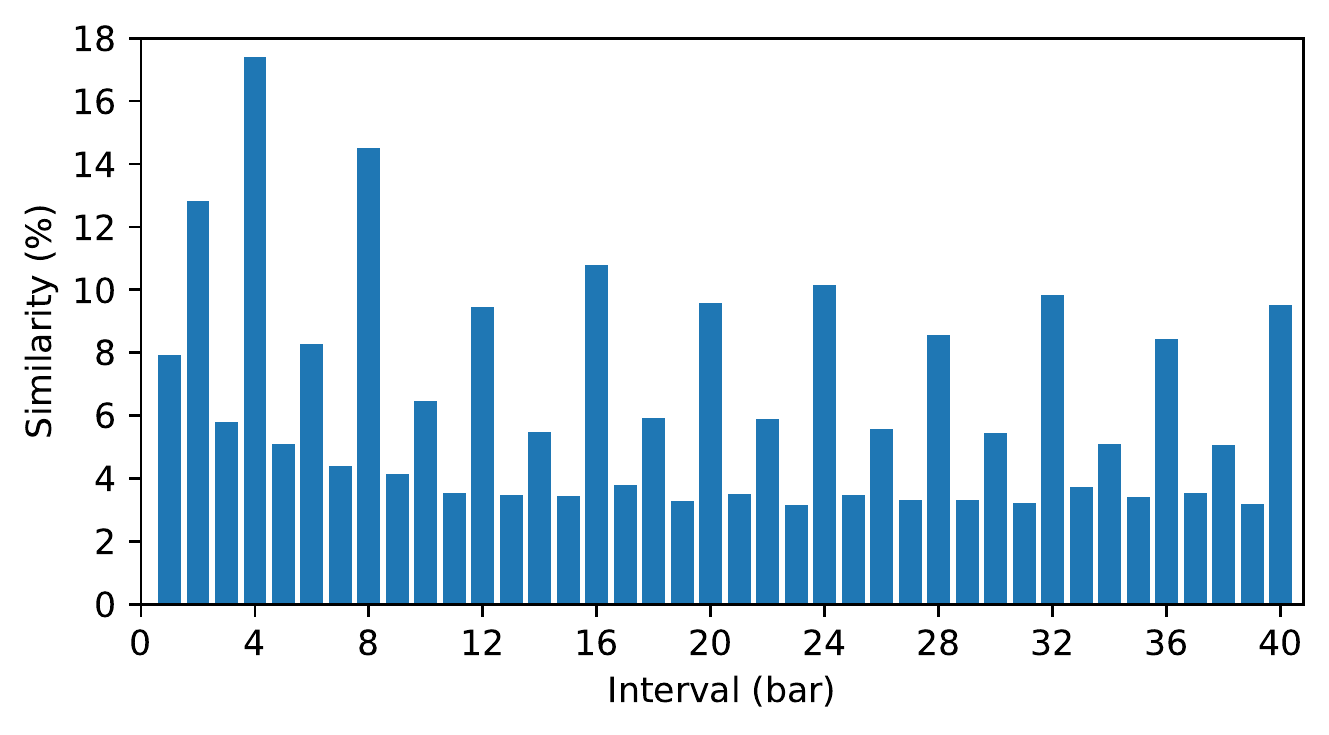}
            % \fbox{\rule[-.5cm]{0cm}{4cm} \rule[-.5cm]{4cm}{0cm}}
            \caption{Rap.}
        \end{subfigure}
        \hfill
        \begin{subfigure}[b]{0.32\textwidth}
            \centering
            \includegraphics[width=\textwidth]{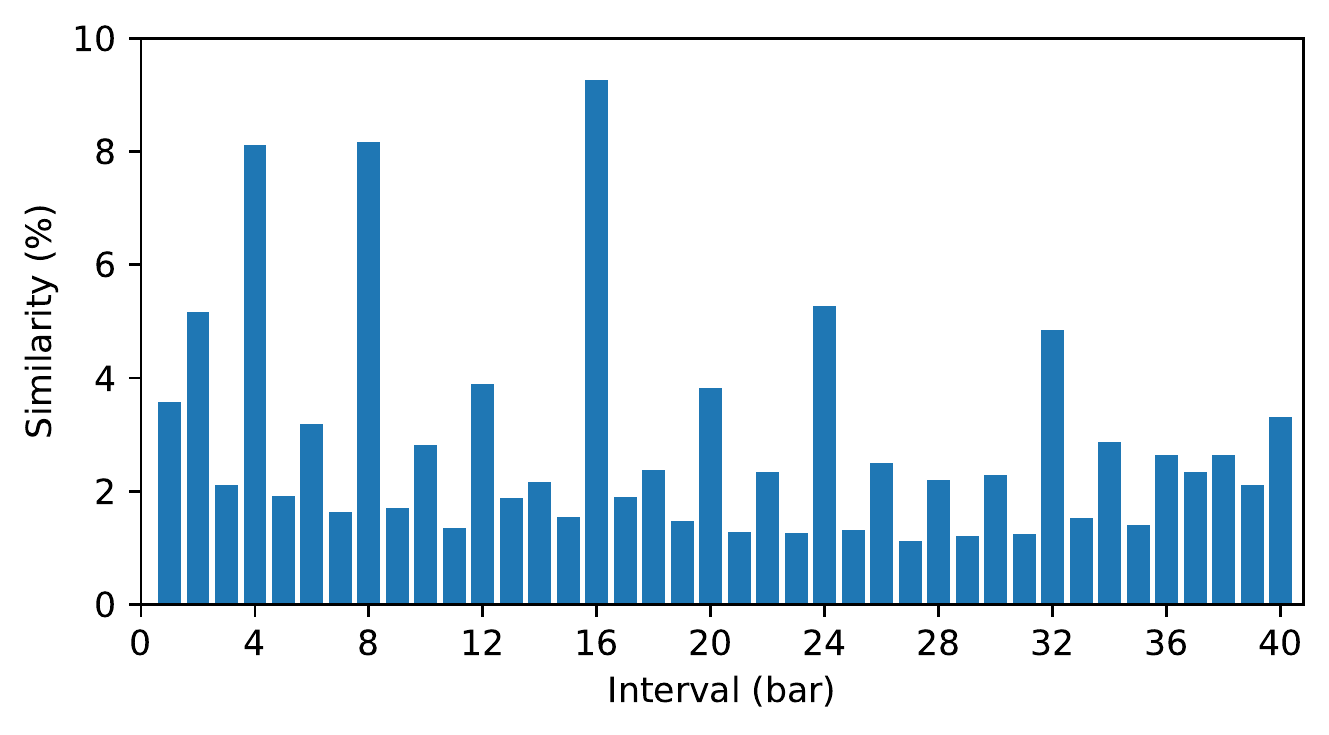}
            % \fbox{\rule[-.5cm]{0cm}{4cm} \rule[-.5cm]{4cm}{0cm}}
            \caption{Jazz.}
        \end{subfigure}
        \hfill
        \begin{subfigure}[b]{0.32\textwidth}
            \centering
            \includegraphics[width=\textwidth]{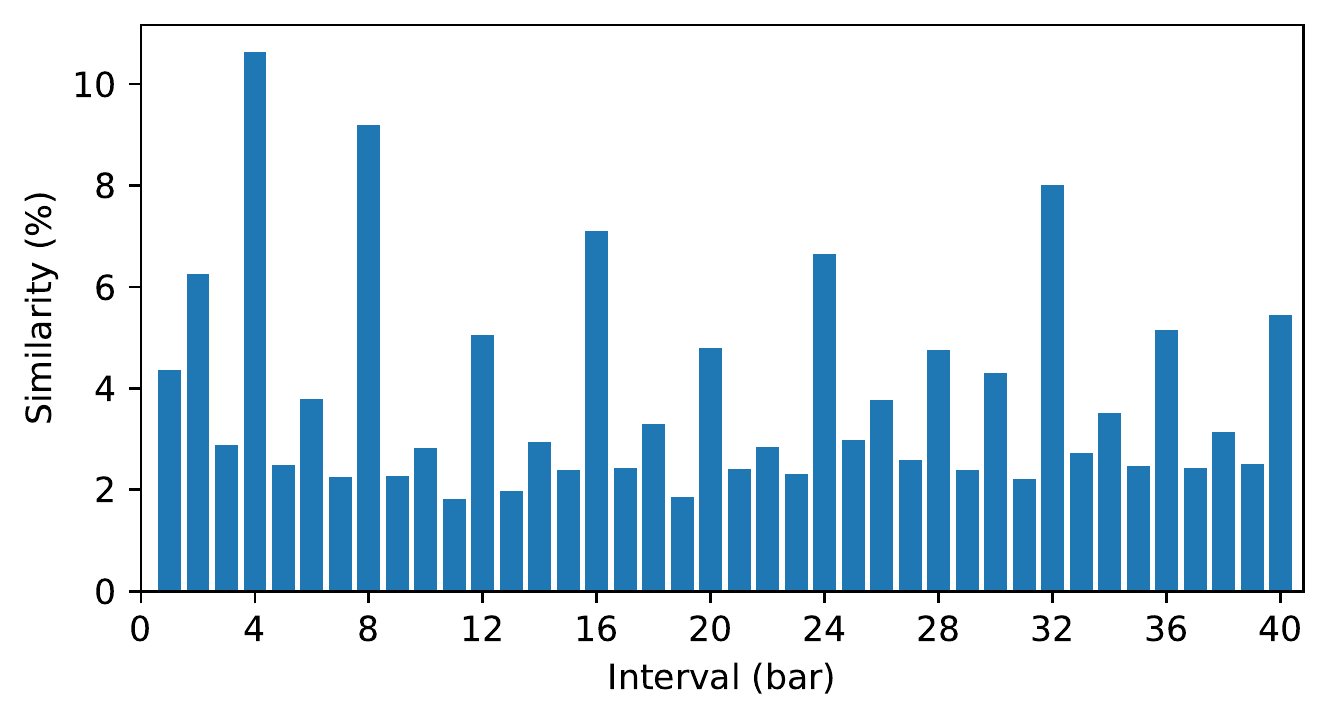}
            % \fbox{\rule[-.5cm]{0cm}{4cm} \rule[-.5cm]{4cm}{0cm}}
            \caption{Latin.}
        \end{subfigure}
        
        \begin{subfigure}[b]{0.32\textwidth}
            \centering
            \includegraphics[width=\textwidth]{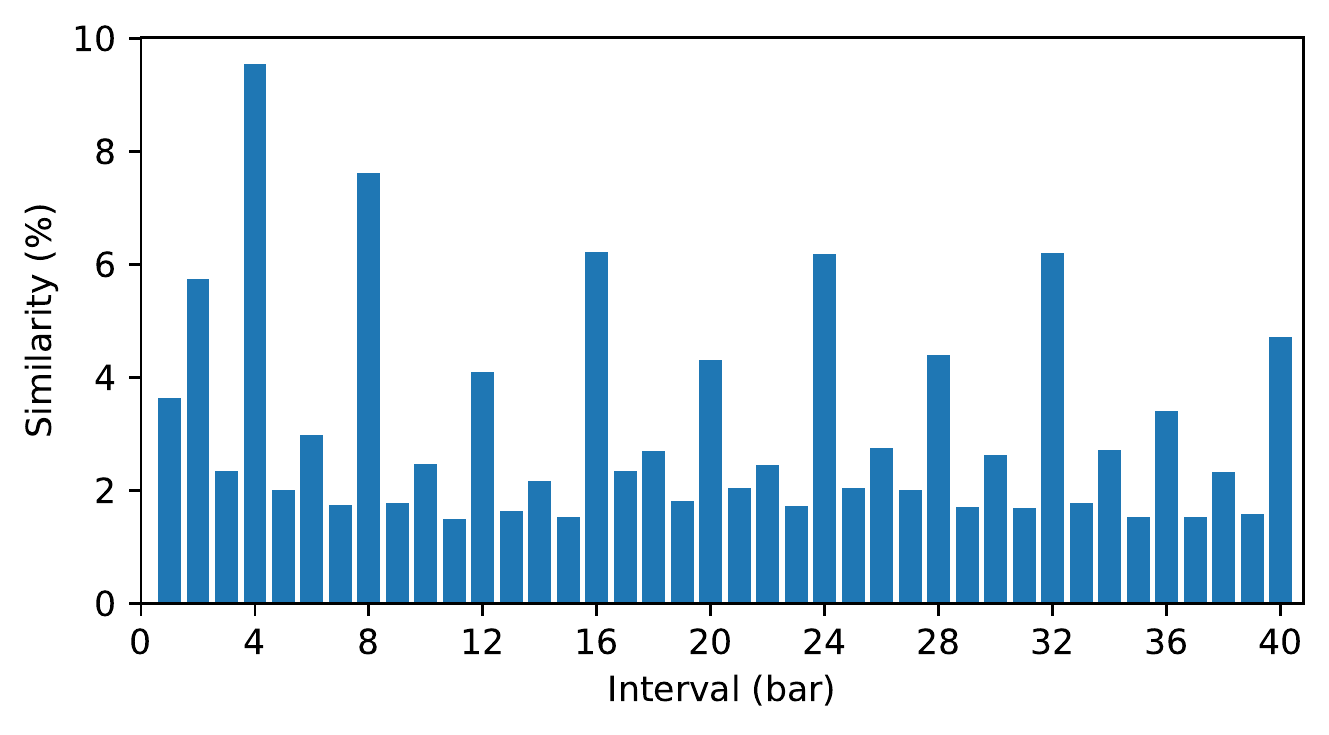}
            % \fbox{\rule[-.5cm]{0cm}{4cm} \rule[-.5cm]{4cm}{0cm}}
            \caption{R\&B.}
        \end{subfigure}
	    \hfill
    	\begin{subfigure}[b]{0.32\textwidth}
            \centering
            \includegraphics[width=\textwidth]{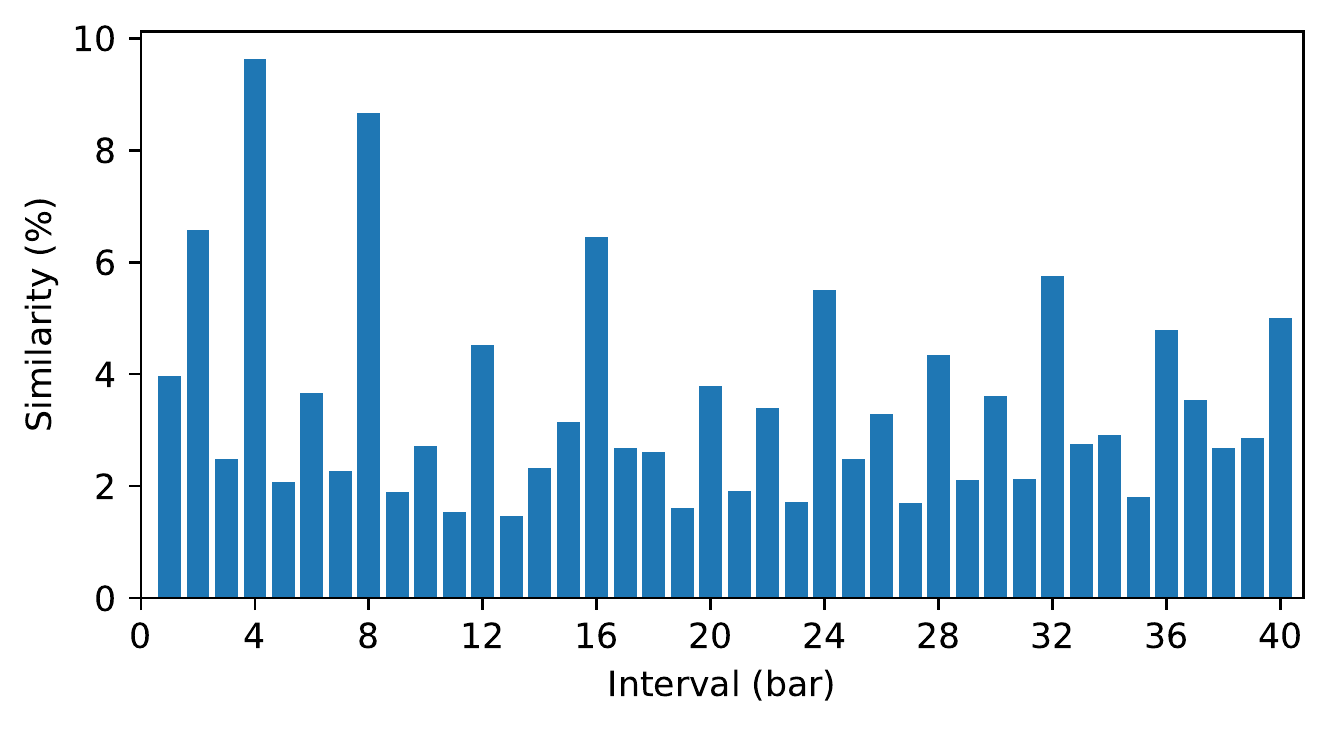}
            % \fbox{\rule[-.5cm]{0cm}{4cm} \rule[-.5cm]{4cm}{0cm}}
            \caption{International.}
        \end{subfigure}
        \hfill
        \begin{subfigure}[b]{0.32\textwidth}
            \centering
            \includegraphics[width=\textwidth]{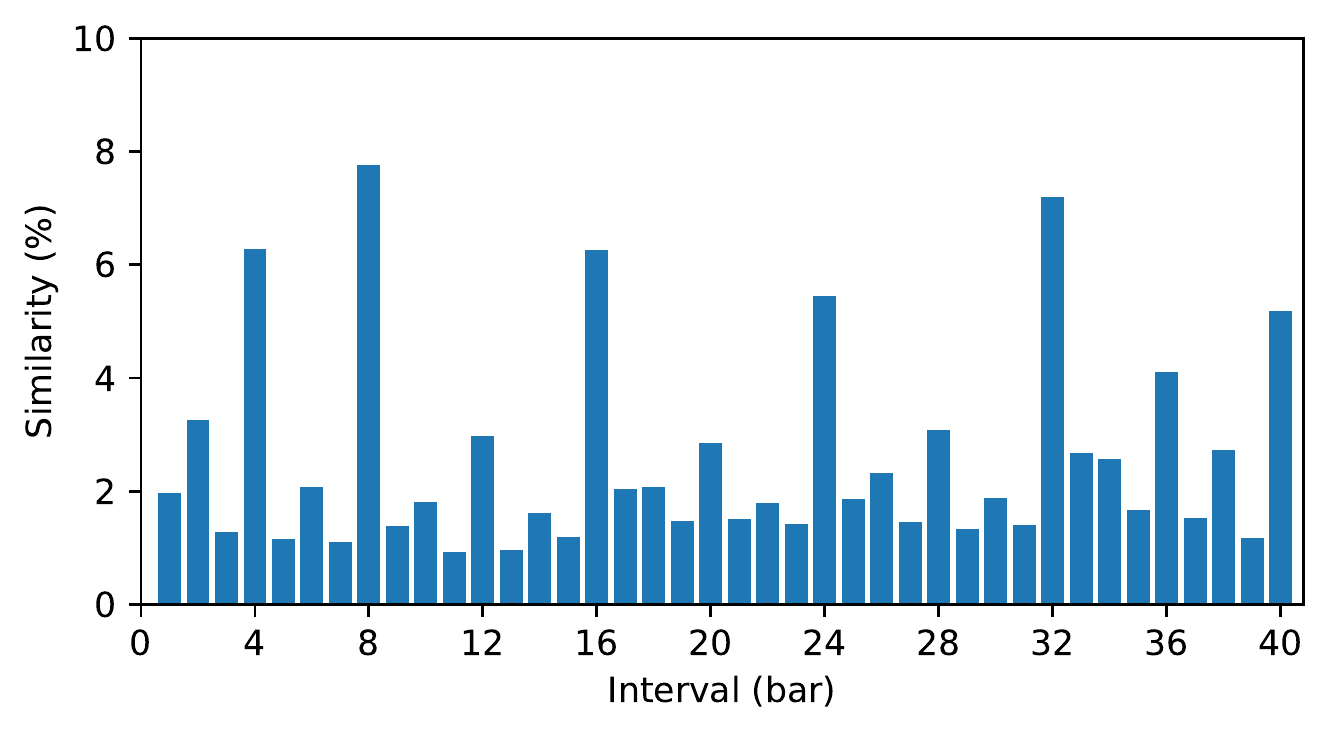}
            % \fbox{\rule[-.5cm]{0cm}{4cm} \rule[-.5cm]{4cm}{0cm}}
            \caption{Country.}
        \end{subfigure}
        
        \begin{subfigure}[b]{0.32\textwidth}
            \centering
            \includegraphics[width=\textwidth]{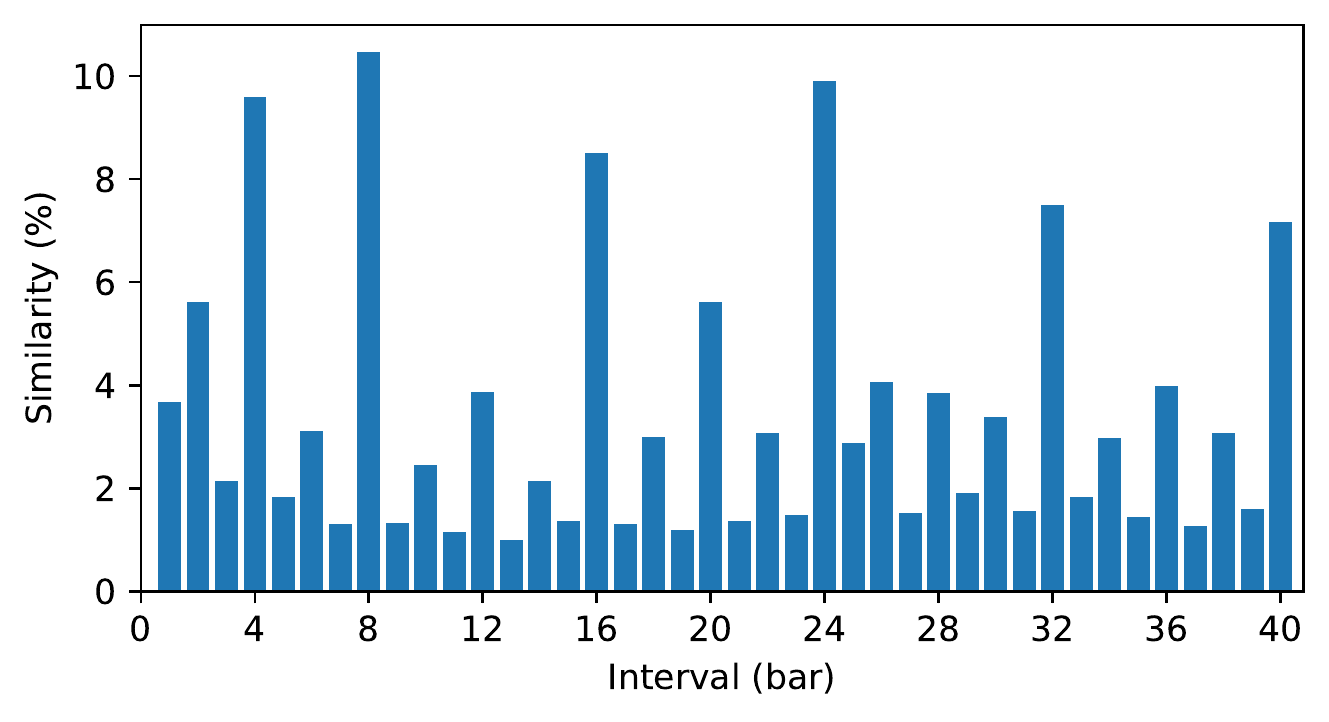}
            % \fbox{\rule[-.5cm]{0cm}{4cm} \rule[-.5cm]{4cm}{0cm}}
            \caption{Reggae.}
        \end{subfigure}
        \hfill
        \begin{subfigure}[b]{0.32\textwidth}
                \centering
                \includegraphics[width=\textwidth]{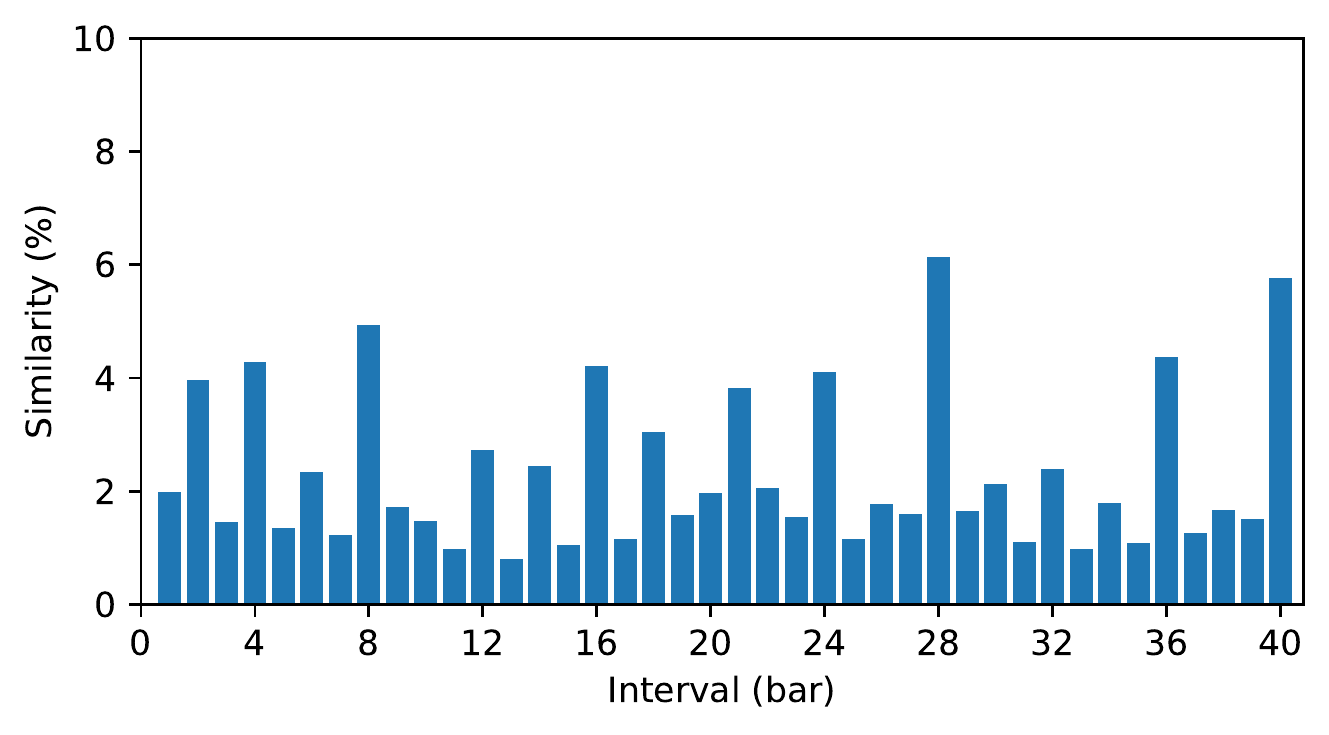}
                % \fbox{\rule[-.5cm]{0cm}{4cm} \rule[-.5cm]{4cm}{0cm}}
                \caption{Blues.}
        \end{subfigure}
        \hfill
        \begin{subfigure}[b]{0.32\textwidth}
                \centering
                \includegraphics[width=\textwidth]{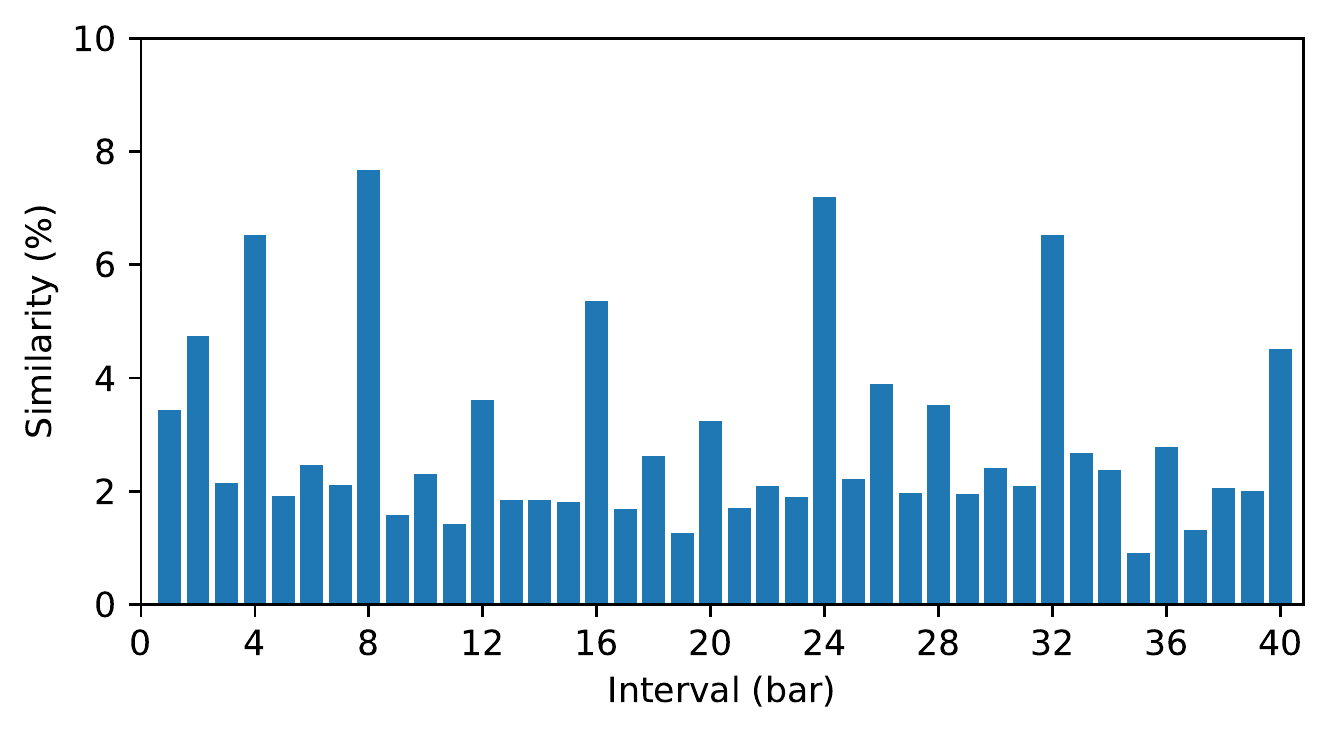}
                % \fbox{\rule[-.5cm]{0cm}{4cm} \rule[-.5cm]{4cm}{0cm}}
                \caption{Vocal.}
            \end{subfigure}
        
        \begin{subfigure}[b]{0.32\textwidth}
            \centering
            \includegraphics[width=\textwidth]{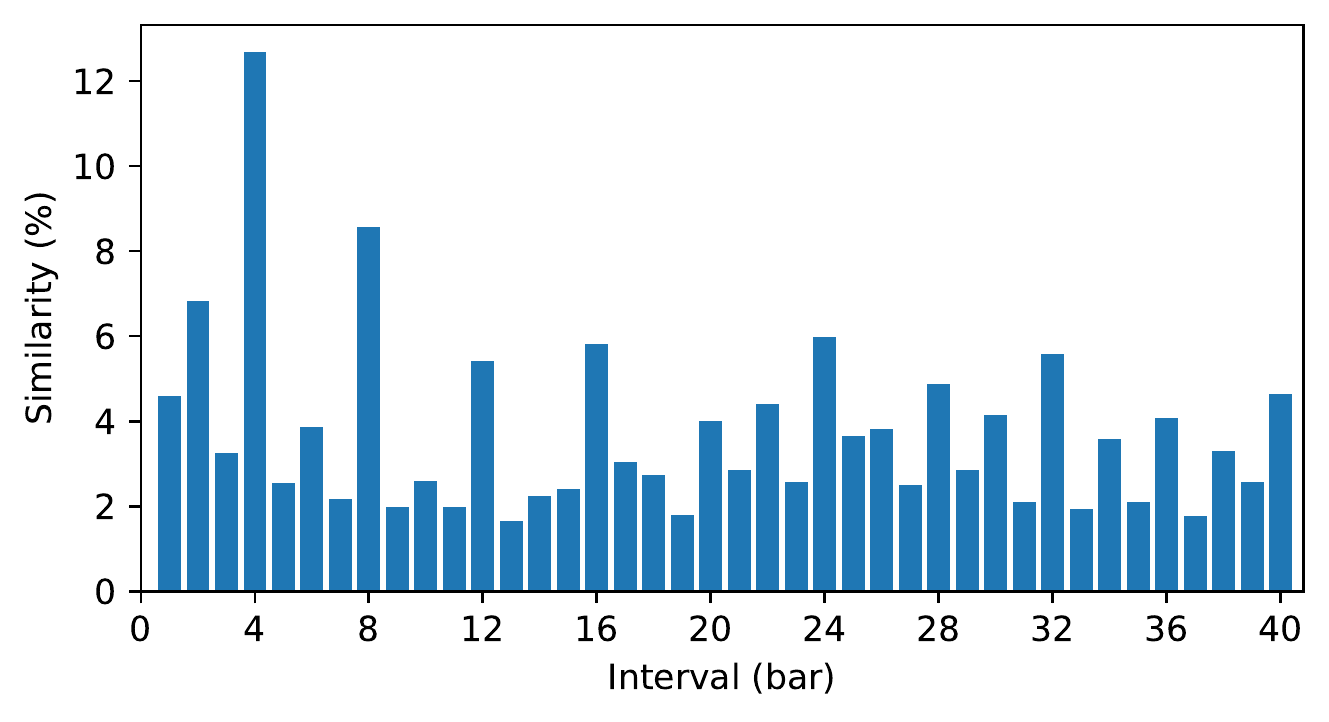}
            % \fbox{\rule[-.5cm]{0cm}{4cm} \rule[-.5cm]{4cm}{0cm}}
            \caption{Folk.}
        \end{subfigure}
        \hfill
        \begin{subfigure}[b]{0.32\textwidth}
            \centering
            \includegraphics[width=\textwidth]{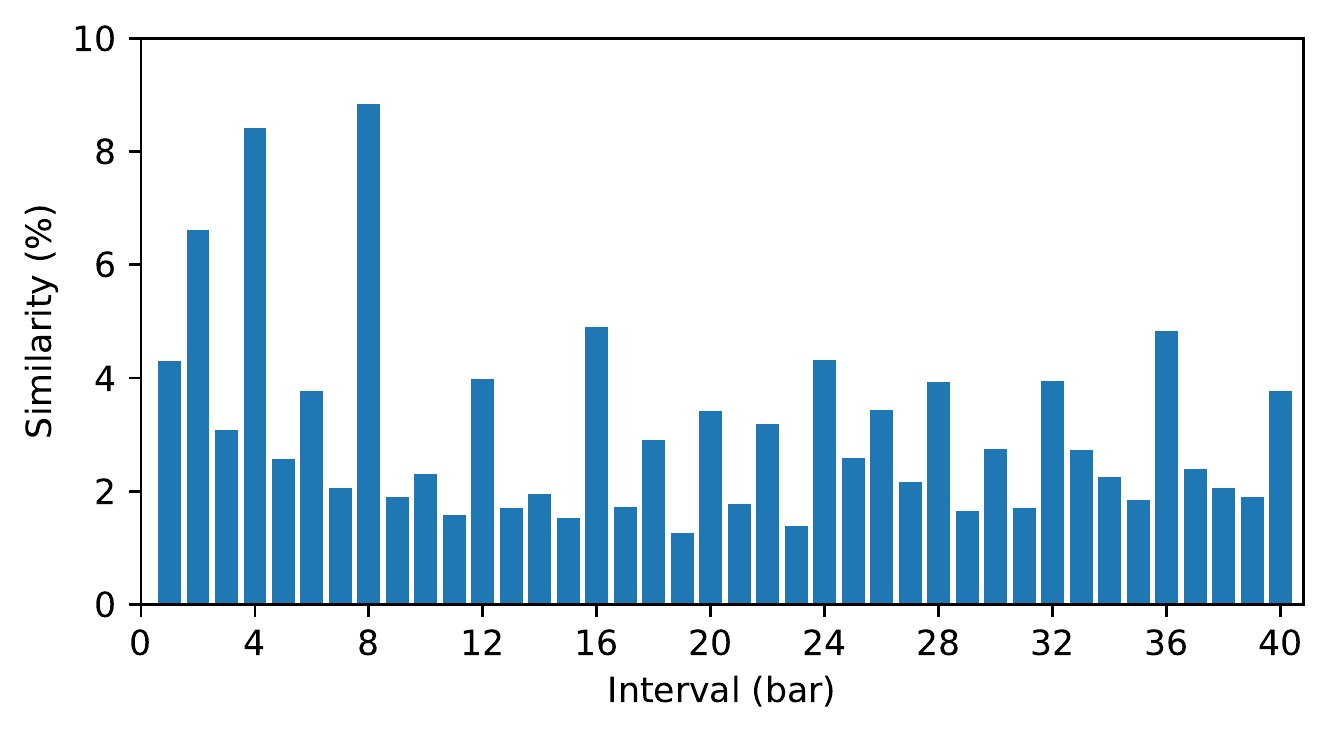}
            % \fbox{\rule[-.5cm]{0cm}{4cm} \rule[-.5cm]{4cm}{0cm}}
            \caption{New age.}
        \end{subfigure}
	\caption{The similarity distribution of the melody track of the different genres in TopMAGD.}
	\label{fig:ds_topmagd}
	\end{figure}

	To see the structure pattern of music of different genres, we conduct the same similarity statistics on the Top-MAGD dataset \footnote{\url{http://www.ifs.tuwien.ac.at/mir/msd/TopMAGD.html}}, which annotates altogether $13$ music genres to the songs. \autoref{fig:ds_topmagd} shows that although different genres have their specific distributions, the general pattern is still applicable to all of these genres.
	
\begin{figure}[htbp]
		\centering
        \includegraphics[width=\textwidth]{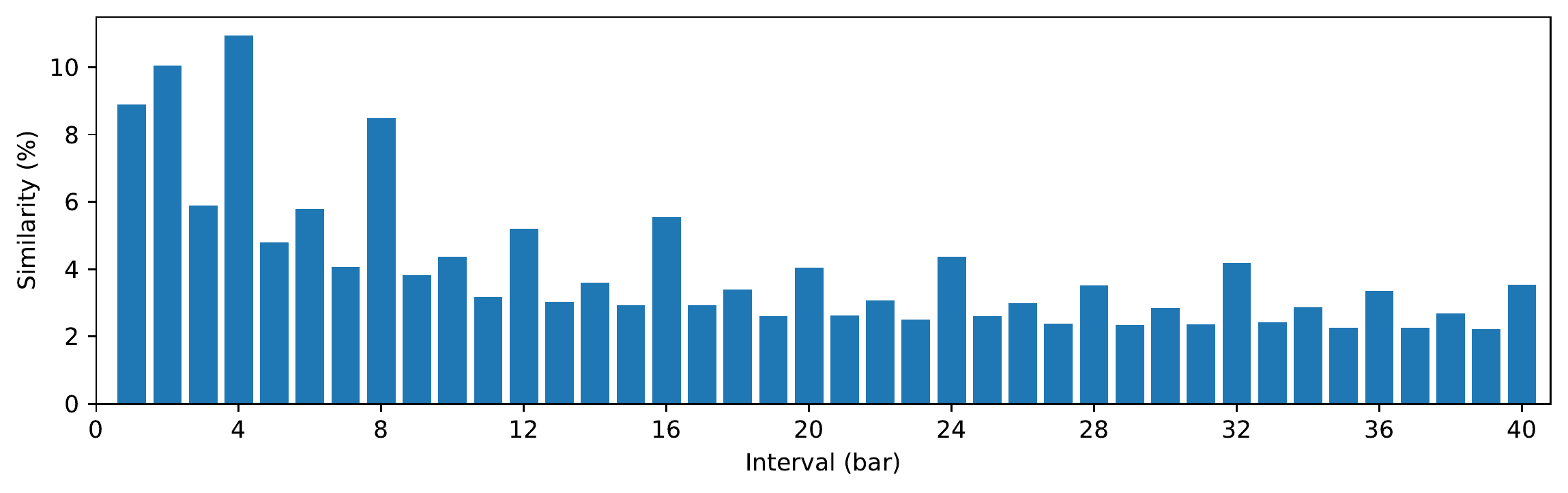}
		\caption{The similarity distribution of all tracks of the Symphony dataset.}
		\label{fig:ds_symphony}
	\end{figure}

	To see whether the pattern still holds on other styles of music, we conduct the statistics on the Symphony dataset \cite{liu2022symphony} and exhibit the distribution in \autoref{fig:ds_symphony}. Since it is not easy to tell the melody tracks for the symphony music, the reported result is calculated over all the tracks. Due to the existence of some instruments that play more repetitions, e.g., drums, the differences among the similarities over the intervals are not that significant, but it still presents the same tendency.
	
    % In addition to the similarity statistics, we further leverage the POP909 dataset \cite{wang2020POP909} where the music structures of each song are manually annotated, to verify the pattern that we have introduced. Specifically ...
	
% 	\begin{figure}[htbp]
% 		\centering
%         \includegraphics[width=\textwidth]{fig/repetition_statistics.png}
% 		\caption{The repetition pattern of the POP909 dataset.}
% 		\label{fig:rp_POP909}
% 	\end{figure}

\section{Details of Experiment Settings}
	\label{sec:app_exp_settings}
	
    \subsection{Dataset Construction}
    \label{subsec:app_exp_dataset}
	
	We use the LMD dataset \cite{raffel2016learning} in our experiments, and perform the following cleaning and processing to ensure the data quality:
	\begin{itemize}[leftmargin=*]
	    \item \textbf{Track Compression}: We first compress various tracks into $6$ tracks \cite{ren2020popmag}, namely square synthesizer, piano, guitar, string, bass, and drum, with the square synthesizer playing the melody.
	    \item \textbf{Note Position and Duration Normalization}: To ensure that the bar splitting is correct and the duration of notes recorded in MIDI files conforms to the musically perfect duration (e.g., quarter notes held for exact $1$ beat, $8$-th notes held for exact $0.5$ beat), we use MuseScore\footnote{\url{https://musescore.org/}} to normalize the note position and duration.
	    \item \textbf{Data Filtering}: Since this dataset is crawled from the Internet and contains many samples of low quality, we use a set of heuristic rules presented in \autoref{tab:data_filtering} to filter them out and keep the samples of good quality and reasonable lengths. 
	    \item \textbf{Pitch Normalization}: We normalize the pitches to transfer the tonality to ``C major'' or ``A minor''.
	\end{itemize}
	
	\begin{table}[ht]
	    \caption{Our filtering rules.}
		\label{tab:data_filtering}
		\begin{center}
			\begin{tabular}{p{1.5cm}p{9cm}p{2cm}}
				\toprule
				Type & Rule & Purpose \\
				\midrule
				Duplication & Remove the duplicated samples that have the same duration and the same numbers of bars, notes, distinct note positions, and instruments. & To remove the duplicated samples. \\
				\midrule
				\multirow[t]{6}{1.5cm}{Musical features} & Only keep the samples of time signature 4/4.  & \multirow[t]{6}{2cm}{To remove musically complicated or erroneous samples.} \\
				                 & Only keep the samples that have at least 2 instruments and have the square synthesizer (the melody track). & \\
				                 & Only keep the samples whose tempo values (the performance speed) are not less than 24 and not larger than 200. & \\
				                 & Only keep the samples whose pitch values are not less than 21 (A0) and not larger than 108 (C8). & \\
				                 & Only keep the samples with maximum note duration not longer than 16 beats (4 bar). & \\
				                 & Remove the samples that contain 4 or more empty bars, or the pitch/duration values are the same for all the notes. & \\
				\bottomrule
			\end{tabular}
		\end{center}
	\end{table}
	
\begin{figure}
		\centering
		\begin{subfigure}[t]{0.49\textwidth}
		    \includegraphics[width=\textwidth]{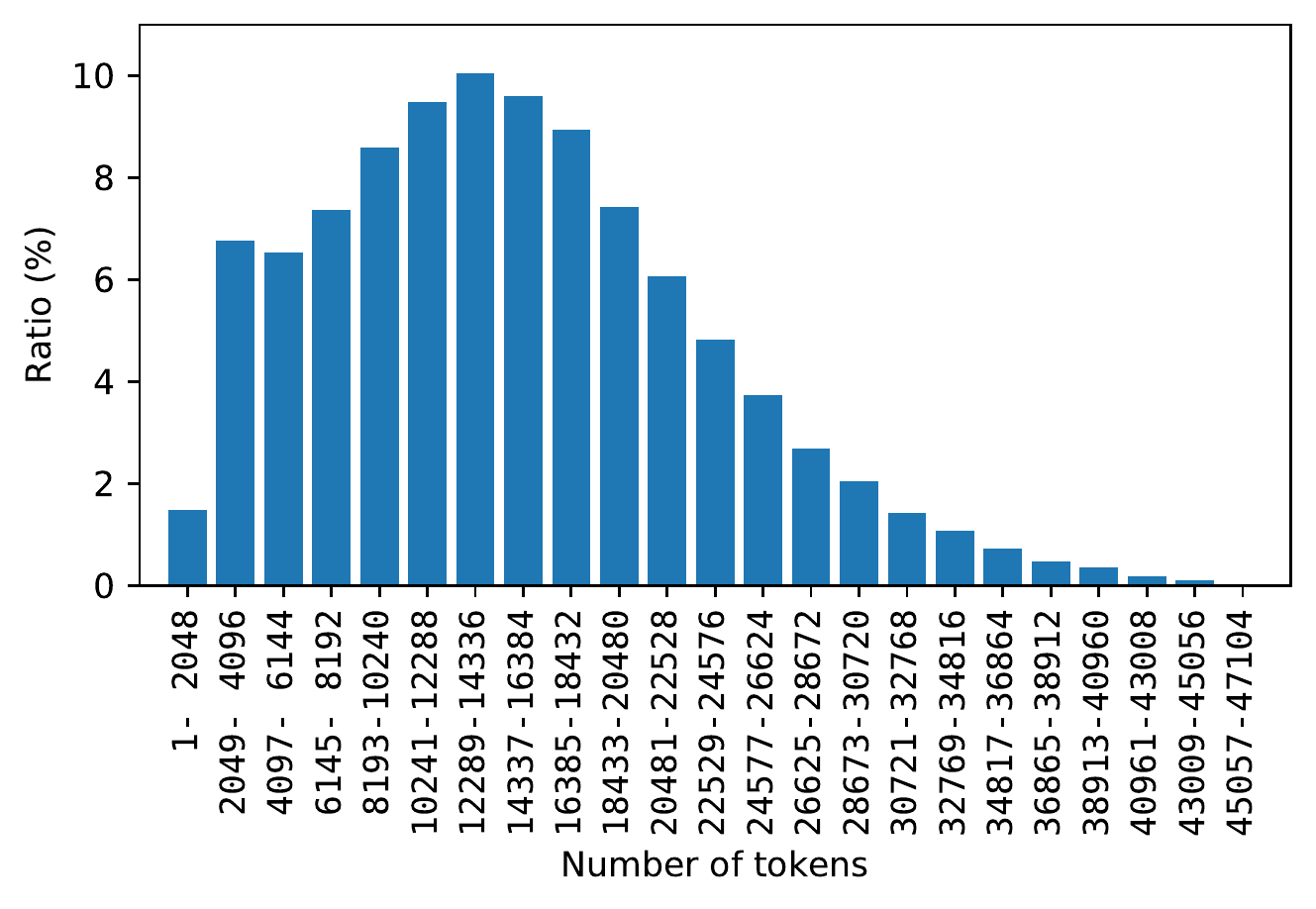}
		    \caption{Statistics of number of tokens.}
		\end{subfigure}
		\hfill
		\begin{subfigure}[t]{0.49\textwidth}
		    \includegraphics[width=\textwidth]{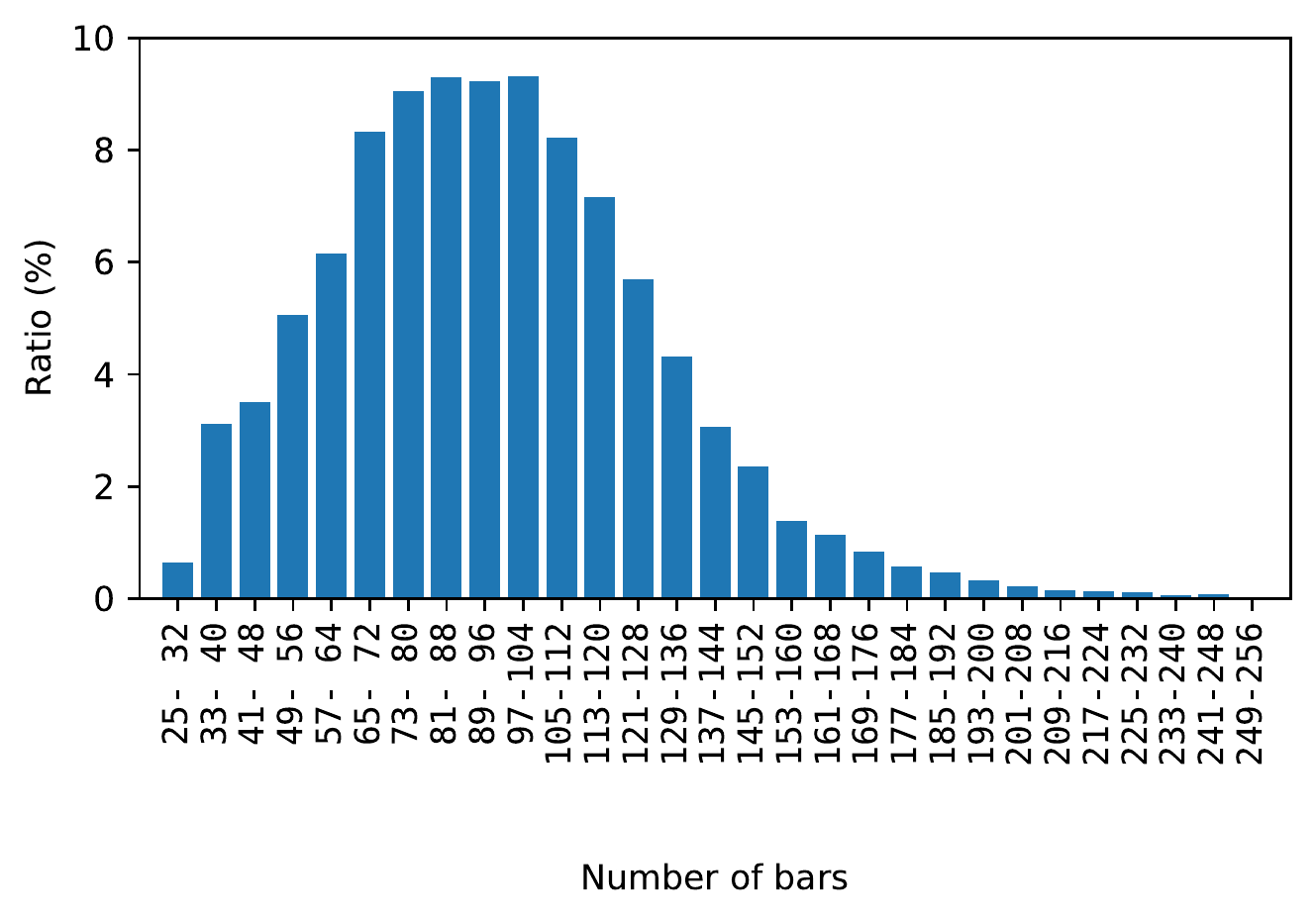}
		    \caption{Statistics of number of bars.}
		\end{subfigure}
		{\caption{Length statistics. The vertical axes represent the ratios of those samples that are in the corresponding ranges.}\label{fig:length_statistics}}
	\end{figure}

	We then represent each MIDI file into a sequence of tokens using a REMI-like \cite{huang2020pop} method. The bar lines are inferred automatically based on the time signature and the note onset positions. The statistics of the number of tokens and the number of bars are shown in \autoref{fig:length_statistics}. The average number of tokens is \num{15042}, the average number of bars is \num{95}, and the average number of tokens per bar is \num{158}. Most samples are longer than \num{10000} tokens, making it essential to design Museformer for efficiently modeling the long music sequences.
	
% 	\begin{figure}[hbpt]
% 		\centering
% 		\floatsetup{valign=t, heightadjust=all}
% 		\ffigbox{
% 		    \begin{subfloatrow}
% 		        \centering
% 	            \ffigbox{\includegraphics[width=0.5\textwidth]{fig/token_num.pdf}}{\caption{Statistics of number of tokens.}}
% 	            \ffigbox{\includegraphics[width=0.5\textwidth]{fig/bar_num.pdf}}{\caption{Statistics of number of bars.}}
% 		    \end{subfloatrow}
% 		}
% 		{\caption{Length statistics. The vertical axes represent the ratios of those samples that are in the corresponding ranges.}\label{fig:length_statistics}}
% 	\end{figure}
	
	\subsection{Implementation Details}
	\label{sec:app_impl}
	
\begin{figure}
		\centering
        \includegraphics[width=0.8\textwidth]{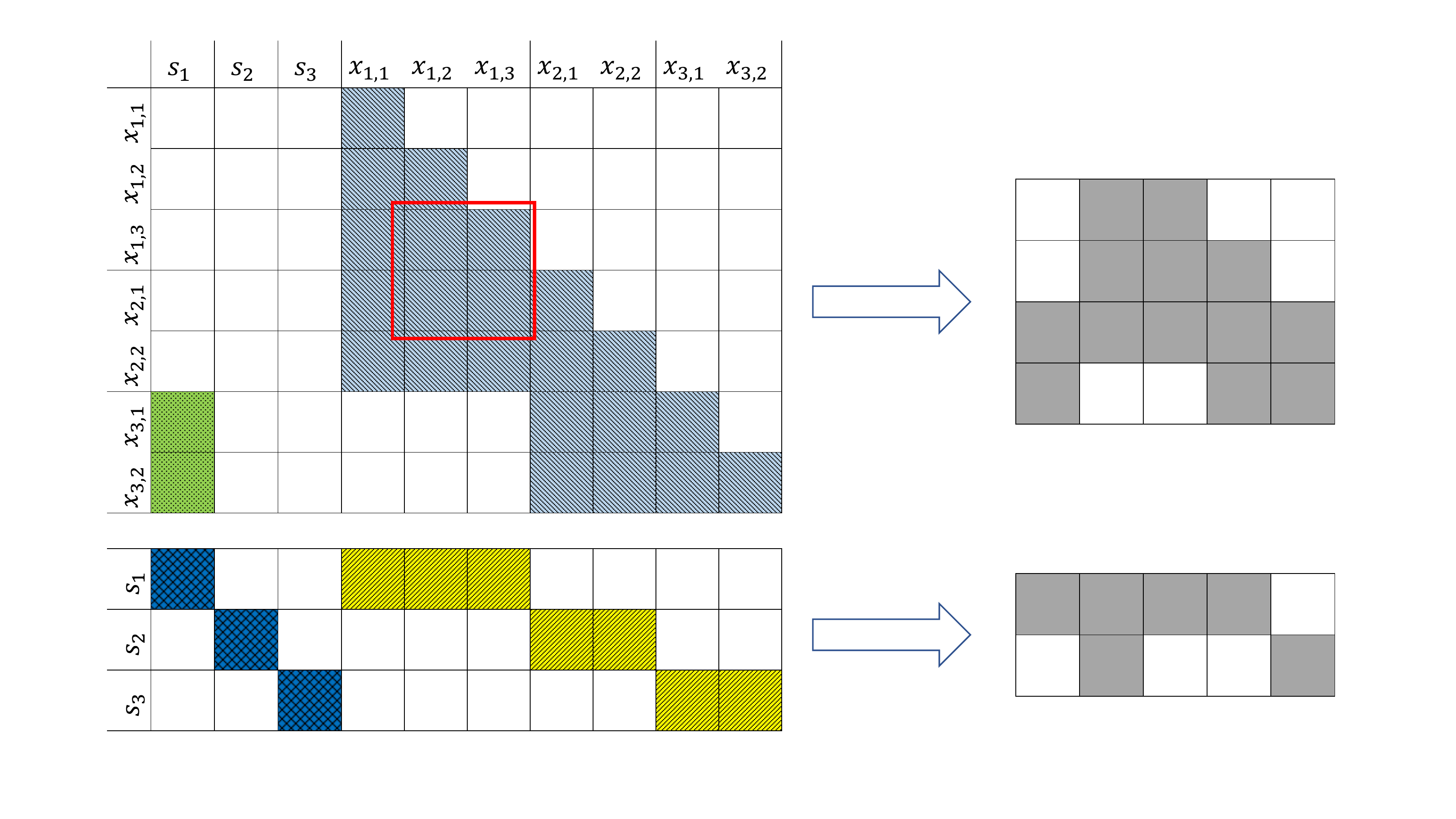}
		\caption{Visualization of the blocksparse implementation. The full attention layouts are on the left, which are exactly the same layouts shown in \autoref{subfig:fc_attn_matrix} except that the summary tokens are put before the music tokens to facilitate our implementation. If the block size is $2$, the corresponding blocksparse layouts are shown on the right. Note that it is only a toy example, and in real cases, the sequences are much longer, and the sparsity of the layouts is much larger than what is depicted here.}
		\label{fig:two_attention_layouts}
	\end{figure}

	Museformer is equipped with a dynamic sparse attention mechanism, which requires constructing distinct attention layouts for each sample according to the token ranges of bars. Since the dynamic sparse attention layout is not regular as sliding windows \cite{beltagy2020longformer,zaheer2020big} or fixed-length blocks \cite{child2019generating}, it is very challenging to leverage the GPU parallel computation techniques to speed up training and decrease the memory consumption. Note that we cannot pad each bar into a fixed-length sequence because the lengths of bars vary drastically, and it would introduce a lot of padding tokens that lead to unacceptable sequence lengths. To achieve an efficient implementation, we utilize a blocksparse method that splits the attention layouts into fixed-size square blocks, and only computes those blocks where there is at least one query-key/value pair that expects computation. In our implementation, all the summary tokens of the input bars are put before the music tokens to facilitate the computation, and thus the summarization step and the aggregation step in FC-Attention can be transferred into computing \textit{summary-to-all} (summary tokens attending to summary tokens and music tokens) and \textit{music-to-all} (music tokens attending to summary tokens and music tokens) attention, respectively, and their attention layouts are shown in \autoref{fig:two_attention_layouts}. We take the following 3 steps to compute each of the two attention processes: 1) \textit{attention layout generation} to generate the full attention layout according to the bar ranges; 2) \textit{layout block-sparsification} to transfer the full layout into a blocksprase layout; 3) \textit{blocksparse computation} to compute the blocksparse multiplications and softmax operations.
	
	\paragraph{Attention Layout Generation}
	According to the bar splitting of each sample, we fill ``true'' on the corresponding areas of a Boolean tensor to construct the attention layout (shown on the left part of \autoref{fig:two_attention_layouts}). To speed up the construction, we collect the begin and end indices of the bars ahead of time and write a CUDA kernel to fill it for all the bars simultaneously.
	
	\paragraph{Layout Block-Sparsification}
	As visualized in \autoref{fig:two_attention_layouts}, the full attention layouts are block-sparsified with a fixed-size square (the block size set to $32$ in our experiments), and diminished into the blocksparse layouts.
	
	\paragraph{Blocksparse Computation}
	We leverage SparTA \cite{SparTA2022} to compute the blocksparse multiplications and softmax operations of attention. It only computes for the shaded areas on the blocksparse layouts in \autoref{fig:two_attention_layouts}.
	
	Since the attention layouts are the same for all the layers and heads in our setting, we only construct the attention layouts once for each sample, and cache them as well as the SparTA kernels for reuse, which saves a lot of memory and time.
	
	\subsection{Detailed Model and Training Configurations}
	
	For Museformer and all the compared models, the basic model and training configurations introduced in \autoref{subsec:exp_settings} are set the same. For Museformer involves two separate attention processes (summarization and aggregation), to make the parameter size comparable with other models, the projection matrices for projecting the targets (i.e., queries) are shared for the two attention processes, and we add different biases for summary tokens and music tokens respectively. It is the same for projecting the sources (i.e., keys and values), except that we use different projection parameters for $\tilde{\bm{S}}_{N(i)}$ in \autoref{eq:aggregation}. All the models have a comparable amount of trainable parameters: Museformer $16.1$M, Music Transformer $16.6$M, Transformer-XL $13.9$M, Longformer $15.3$M, Linear Transformer $13.2$M, with acceptable differences due to different architectures and implementations.
	
	All the models are trained on 4 Nvidia V100 $32$GB GPUs with fp16. Since in Museformer, a token directly attends to the tokens of $8$ previous bars (the selected structure-related bars) and the current bar via the fine-grained attention, for a fair comparison, we set the window sizes for Longformer to \num{1408}, which is approximately the number of tokens that $9$ bars contain. For Music Transformer that uses full attention and cannot process long sequences at once, we chunk the input sequence into blocks of fixed size \num{1408}, and manipulate the batch size and the update frequency to ensure that it is updated the comparable number of times within each epoch as other models. For Transformer-XL, the chunk size and the memory size are also set to \num{1408}. During inference, all the models are applied to generate sequences not shorter than \num{2048} and not longer than \num{20480}, until the end-of-sentence token is generated. The top-$k$ sampling is used, and $k$ is set to $8$.
	
\section{Similarity Distributions of Generated Music}
\label{sec:generated_similarity_analysis}

    Compared to other models (the compared models and the ablation settings), the music generated by Museformer has the similarity distribution most similar to that of the training data, as its SE reported in \autoref{tab:objective_result} is the smallest.
    However, the value of SE may not comprehensively represent the structural characteristics, so we further display and discuss about the specific distributions in this section.
    
\begin{figure}[htbp]
		\centering
		\begin{subfigure}[b]{\textwidth}
            \centering
            \includegraphics[width=\textwidth]{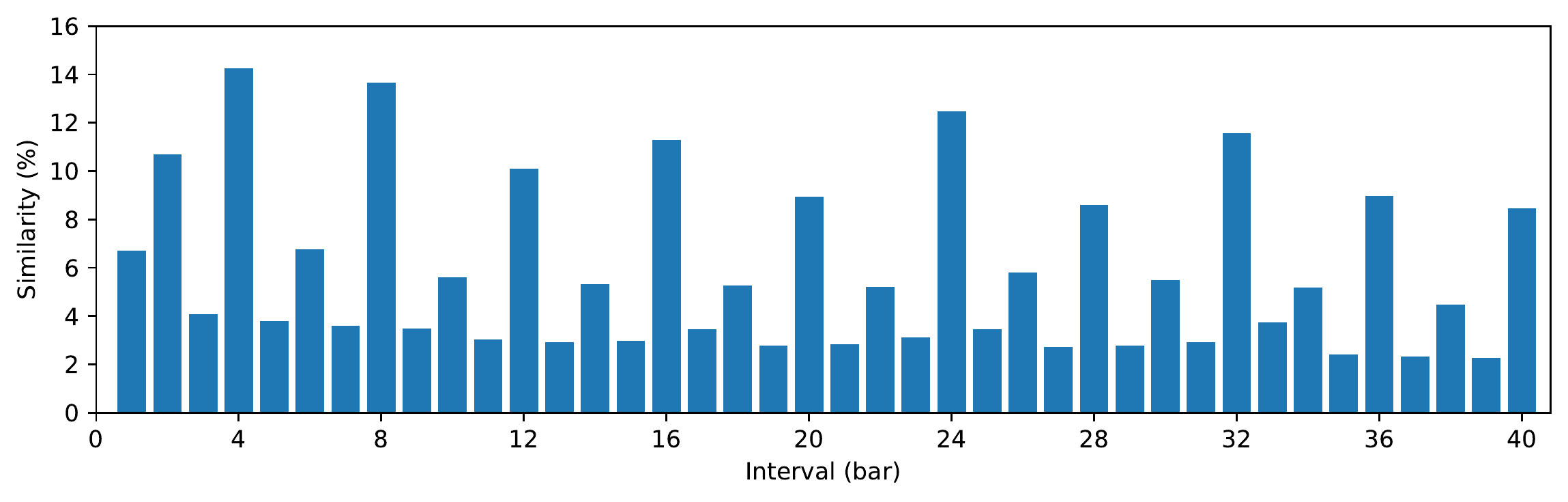}
            \caption{Museformer.}
            \label{fig:gs_mf}
        \end{subfigure}
		
        \begin{subfigure}[b]{0.49\textwidth}
            \centering
            \includegraphics[width=\textwidth]{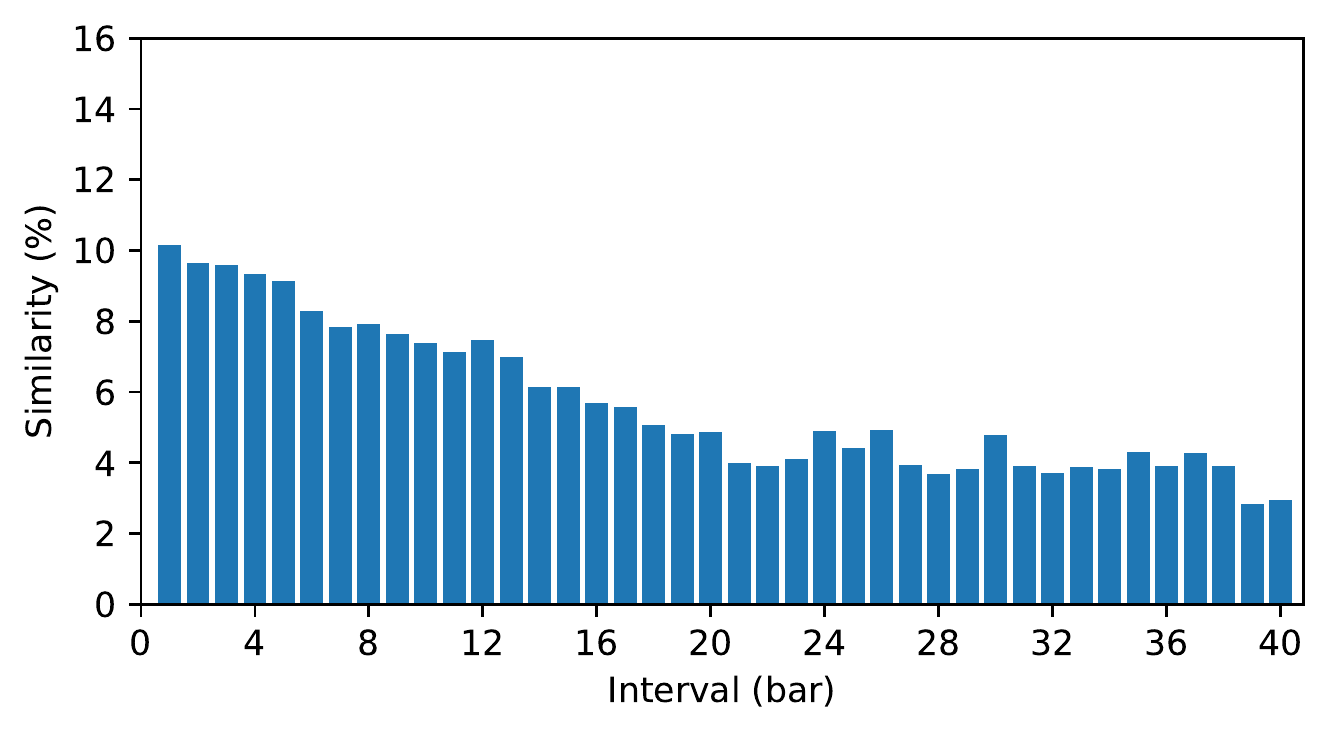}
            \caption{Music Transformer.}
            \label{fig:gs_mt}
        \end{subfigure}
        \hfill
        \begin{subfigure}[b]{0.49\textwidth}
            \centering
            \includegraphics[width=\textwidth]{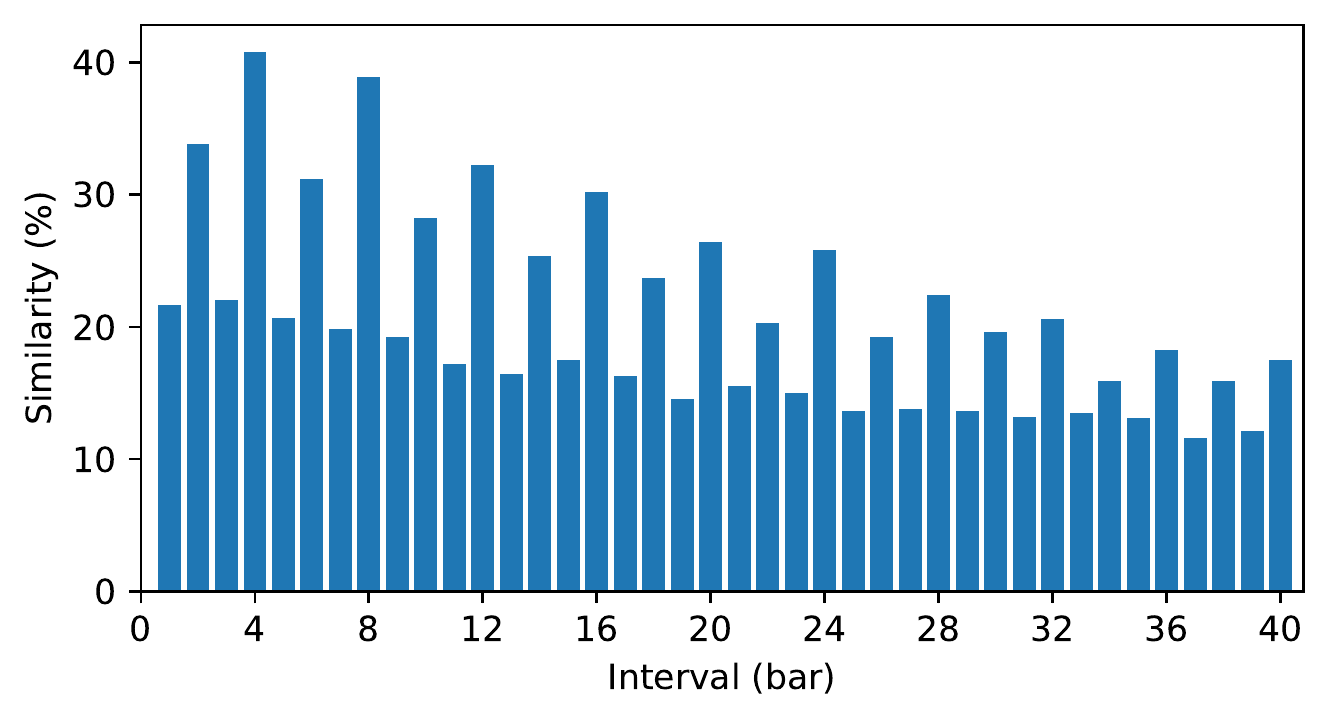}
            % \fbox{\rule[-.5cm]{0cm}{4cm} \rule[-.5cm]{4cm}{0cm}}
            \caption{Transformer-XL.}
            \label{fig:gs_xl}
        \end{subfigure}
        
        \begin{subfigure}[b]{0.49\textwidth}
            \centering
            \includegraphics[width=\textwidth]{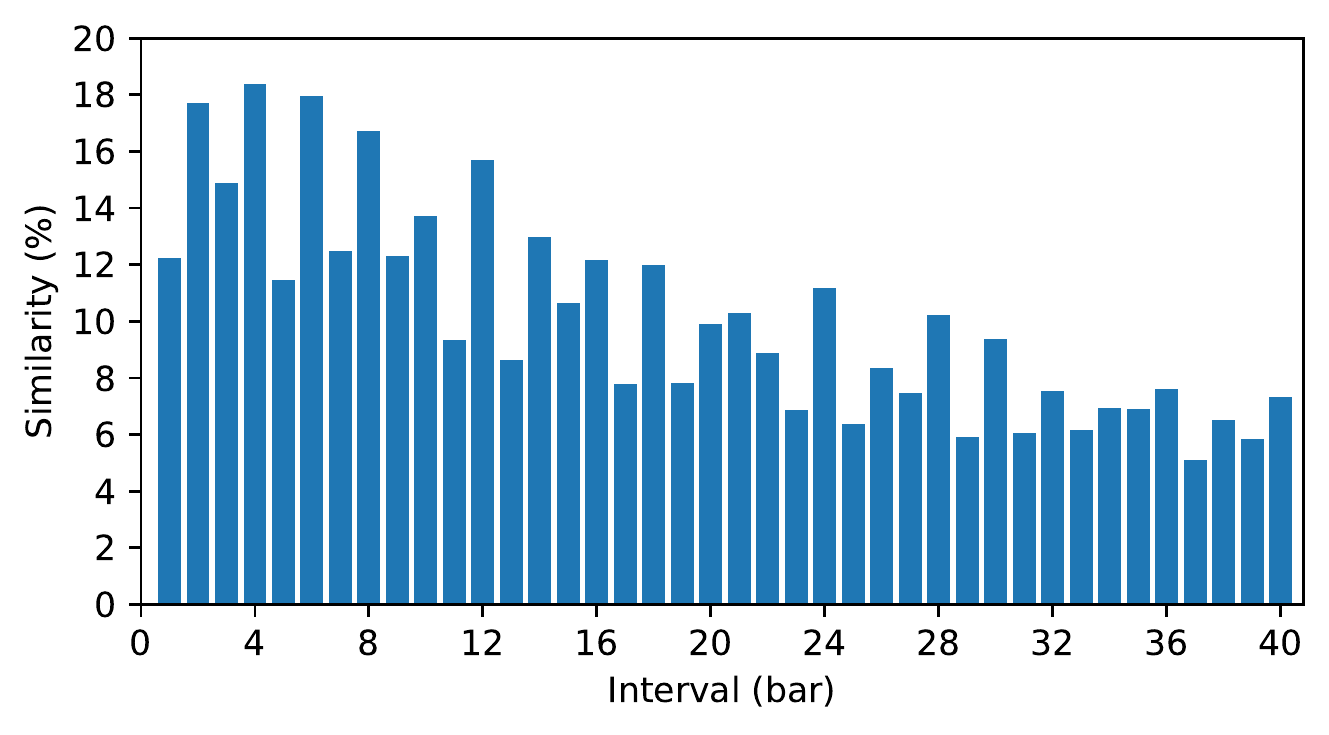}
            % \fbox{\rule[-.5cm]{0cm}{4cm} \rule[-.5cm]{4cm}{0cm}}
            \caption{Longformer.}
            \label{fig:gs_lolm}
        \end{subfigure}
        \hfill
        \begin{subfigure}[b]{0.49\textwidth}
            \centering
            \includegraphics[width=\textwidth]{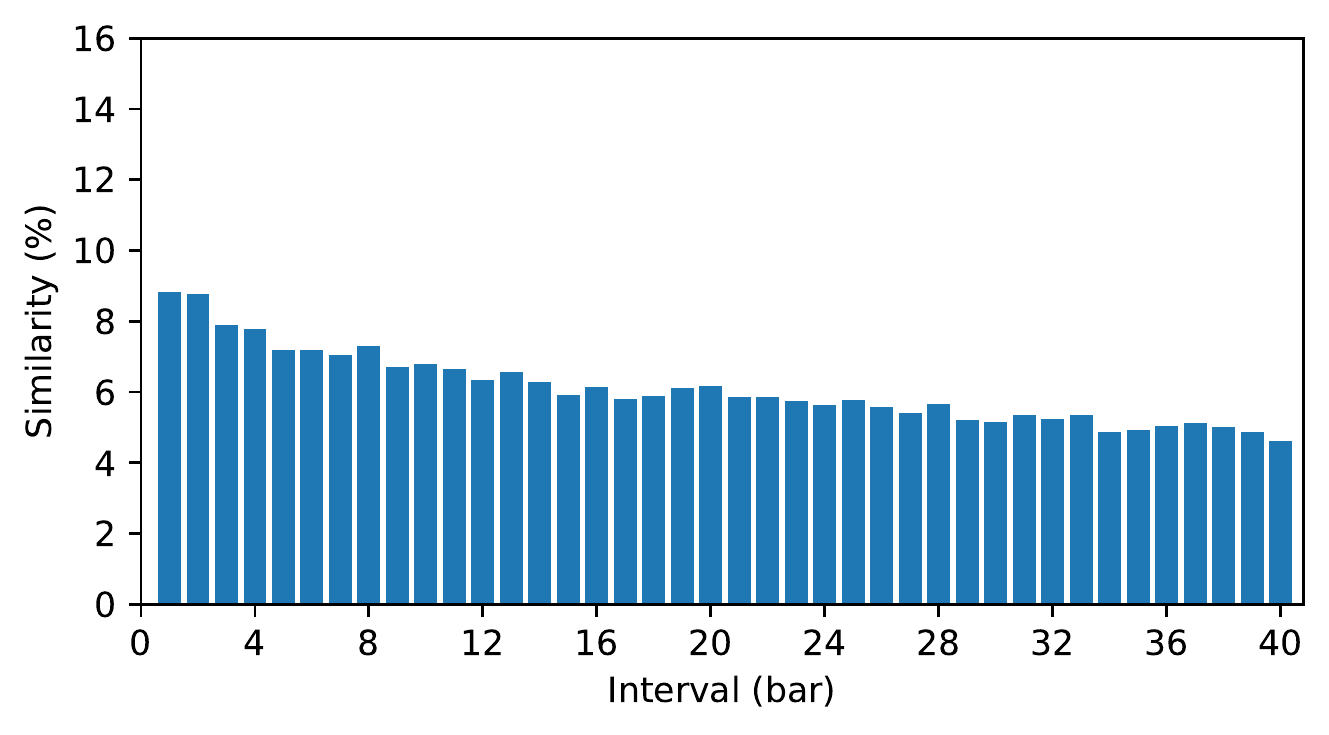}
            % \fbox{\rule[-.5cm]{0cm}{4cm} \rule[-.5cm]{4cm}{0cm}}
            \caption{Linear Transformer.}
            \label{fig:gs_linear}
        \end{subfigure}
        
        \begin{subfigure}[b]{0.49\textwidth}
            \centering
            \includegraphics[width=\textwidth]{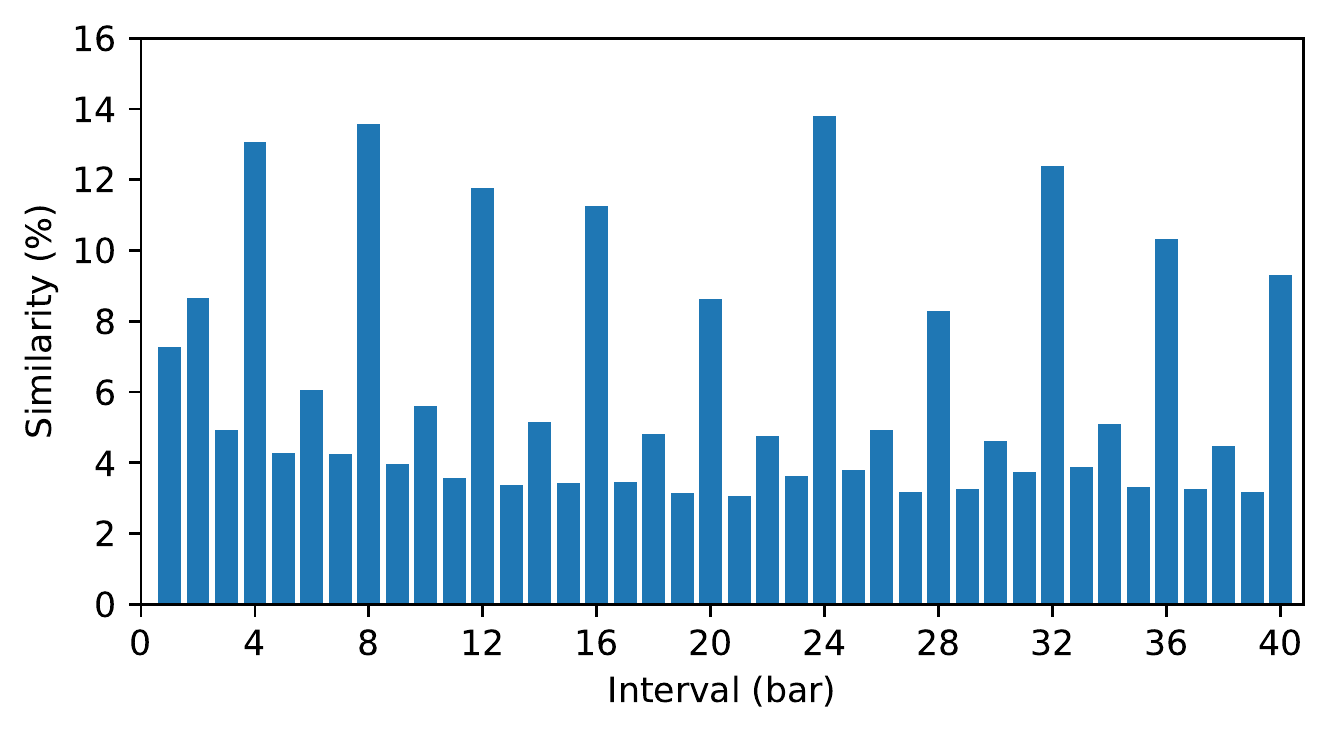}
            % \fbox{\rule[-.5cm]{0cm}{4cm} \rule[-.5cm]{4cm}{0cm}}
            \caption{Museformer w/o coarse-grained.}
            \label{fig:mf_sm}
        \end{subfigure}
        \hfill
        \begin{subfigure}[b]{0.49\textwidth}
            \centering
            \includegraphics[width=\textwidth]{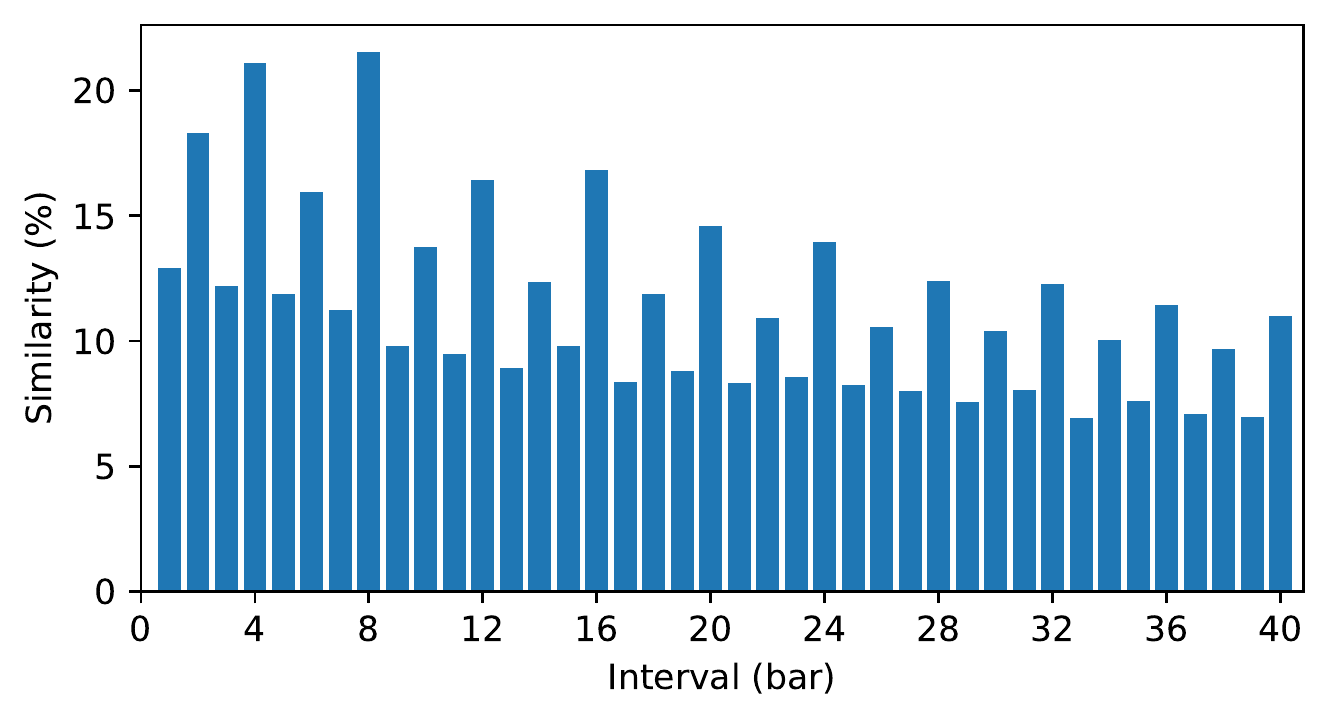}
            % \fbox{\rule[-.5cm]{0cm}{4cm} \rule[-.5cm]{4cm}{0cm}}
            \caption{Museformer w/o bar selection.}
            \label{fig:mf_bs}
        \end{subfigure}
		\caption{The similarity distribution of the melody track of the music generated by different models.}
		\label{fig:gs}
	\end{figure}

    \autoref{fig:gs} shows the similarity distributions. We can observe that:
    1) The distribution of Museformer is very similar to that of the training data (shown in \autoref{fig:similarity_statistics}). The quantity is close, and the contour shows the same periodical pattern, i.e., the previous $2$ bars, the previous $4$-th bar as well as its multipliers, have relatively large similarities.
    2) The distributions of Music Transformer and Linear Transformer show no periodical pattern. It implies that the Music Transformer model trained on short sequences cannot generate well-structured music of long lengths, and Linear Transformer cannot well capture the structure-related correlations even though its receptive field covers the whole sequence.
    3) The distributions of Transformer-XL and Longformer show the tendency of the periodical pattern, and the similarity decreases as the interval increases in general. It indicates that the two models whose receptive fields only contain the most recent content have the ability to generate periodical repetitions of short distances but fall short for long-term structures.
    4) It seems that compared to the quantity of the similarity, the contour (i.e., the periodical pattern) is more relevant to the human scoring on the structure-related metrics. The distributions of Transformer-XL and Longformer show the pattern, and their subjective scores on short-term and long-term structures (shown in \autoref{tab:subjective_results_score}) are relatively high, and are higher than Music Transformer and Linear Transformer whose distributions fail to show the periodical pattern. However, the quantity of the similarity may also influence human decisions. For example, the quantities of the similarities of Transformer-XL are relatively high, which indicates that too many repetitions are generated. For these samples, human scorers sometimes think them annoying and eventually give lower scores on musicality.
    5) The ablation setting, Museformer w/o coarse-grained, has slightly larger SE compared to Museformer, and its distribution clearly shows the periodical pattern. Thus, the coarse-grained attention does not contribute a lot to the music structures. The distribution of Museformer w/o bar selection shows the tendency of the periodical pattern, and the similarity decreases in general as the interval increases, similar to those of Transformer-XL and Longformer. It indicates that the structure-related bars are contributory to generating music with both short-term and long-term structures. 
	
\end{document}